\documentclass[aps,prd,twocolumn,showpacs,superscriptaddress,amsmath]{revtex4-1}

\usepackage{longtable}
\usepackage{graphicx}
\usepackage{amsmath,amssymb}
\usepackage{natbib}
\usepackage{color}
\usepackage{bbm}
\usepackage[colorlinks=true,linkcolor=blue,citecolor=magenta,pdfstartview=FitB,bookmarks=false]{hyperref}

\newcommand{\e}{{\rm e}}

\newcommand{\irm}{{\rm i}}
\newcommand{\beq}{\begin{equation}}
\newcommand{\eeq}{\end{equation}}
\newcommand{\bdm}{\begin{displaymath}}
\newcommand{\edm}{\end{displaymath}}

\newcommand{\gtorder}{\mathrel{\raise.3ex\hbox{$>$}\mkern-14mu
            \lower0.6ex\hbox{$\sim$}}}
\newcommand{\ltorder}{\mathrel{\raise.3ex\hbox{$<$}\mkern-14mu
            \lower0.6ex\hbox{$\sim$}}}

\begin{document}

\title{Low-Frequency Terrestrial Gravitational-Wave Detectors}

\author{Jan Harms}
\affiliation{INFN, Sezione di Firenze, Sesto Fiorentino, Italy}
\author{Bram J. J. Slagmolen}
\affiliation{The Australian National University, Centre for Gravitational Physics, Canberra, Australia}
\author{Rana X. Adhikari}
\affiliation{LIGO Laboratory, California Institute of Technology, Division of Physics, Math, and Astronomy, Pasadena, California}
\author{M. Coleman Miller}
\affiliation{Department of Astronomy and Joint Space-Science Institute, University of Maryland, College Park, MD 20742-2421, USA}
\affiliation{Department of Physics and Astronomy, Johns Hopkins University, Baltimore, MD 21218}
\author{Matthew Evans}
\affiliation{Kavli Institute for Astrophysics, Massachusetts Institute of Technology, Cambridge, Massachusetts}
\author{Yanbei Chen}
\affiliation{California Institute of Technology, Division of Physics, Math, and Astronomy, Pasadena, California}
\author{Holger M\"uller}
\affiliation{Department of Physics, University of California, Berkeley, California}
\author{Masaki Ando}
\affiliation{Department of Physics, the University of Tokyo, Tokyo 113-0033, Japan}
\affiliation{National Astronomical Observatory of Japan, Tokyo 181-8588, Japan}

\date{\today}

\begin{abstract}
Direct detection of gravitational radiation in the audio band is being pursued with a network 
of kilometer-scale interferometers (LIGO, Virgo, KAGRA). Several space missions (LISA, 
DECIGO, BBO) have been proposed to search for sub-Hz radiation from massive astrophysical 
sources. Here we examine the potential sensitivity of three ground-based detector concepts 
aimed at radiation in the 0.1 -- 10\,Hz band. We describe the plethora of potential astrophysical 
sources in this band and make estimates for their event rates and thereby, the sensitivity 
requirements for these detectors. The scientific payoff from measuring astrophysical gravitational 
waves in this frequency band is great. Although we find no fundamental limits to the detector 
sensitivity in this band, the remaining technical limits will be extremely challenging to overcome.
\end{abstract}

\pacs{04.80.Nn, 09.30.Fn, 95.75.Wx, 95.55.Ym, 37.25,+k, 04.30.Tv, 04.30.Db}

\maketitle

\section{Introduction}
\label{sec:intro}
Gravitational waves (GWs) in the context of General Relativity promise to reveal new information about
the bulk motions of massive compact objects in the universe. This new view of the universe will
complement our existing, electro-magnetic understanding. In this decade, kilometer-scale interferometers (such as LIGO \cite{LSC2009b}, Virgo \cite{Virgo:2011}, GEO600 \cite{LuEA2010}, and KAGRA \cite{Som2012}) are expected to make the first direct detections of GWs in the 10 -- 10000\,Hz band~\cite{Rana:Review}. These waves would be associated with the coalescence of neutron-star binaries and low-mass black-hole binaries. In the proposed underground Einstein Telescope, the approach is to improve the traditional detector design to extend the detection band down to 3\,Hz \cite{PuEA2010}. A set of space interferometer missions (eLISA \cite{ESA:LISA}, DECIGO \cite{Ando:DECIGO}, BBO \cite{Phi2003}) have been proposed to search for the gravitational waves from supermassive black holes as well as the inspiral phase of the low-mass compact objects~\cite{eLISA:Science}.

The reason for constructing interferometers in space is chiefly to avoid the seismic disturbances
on the Earth due to natural and anthropogenic sources. Even if we posit a very sophisticated
vibration isolator, a GW detector on the Earth cannot be shielded from the fluctuations in the
terrestrial gravitational forces~\cite{Sau1984,HuTh1998} (a.k.a. 
Newtonian noise or gravity-gradient noise). In this work we argue that it is possible, with
reasonable extrapolations of existing technology, to make detections of GWs in the 0.1 -- 10\,Hz
using terrestrial detectors.

In Section~\ref{sec:atom}, we describe an atom interferometer with improved immunity
to technical noise sources. In Section~\ref{sec:TOBA}, we explore improvements in a 
previously proposed differential torsion bar detector. In Section~\ref{sec:Michelson}, we propose
a version of the standard Michelson interferometer optimized for low frequency sensitivity.
In Section~\ref{sec:Newton} we explore options for mitigating the effects of the Newtonian
gravitational noise. Finally, in Section~\ref{sec:Sources}, we explore what sources of gravitational 
waves can be probed using this set of terrestrial, low-frequency detectors. As will be shown, the 
sensitivity at 0.1\,Hz to GWs should be around $10^{-20}\,\rm Hz^{-1/2}$ or better, and the 
corresponding instrumental designs will be referred to as MANGO in this paper.

\section{Atom Interferometers}
\label{sec:atom}

\begin{figure}[h]
\includegraphics[width=0.9\columnwidth]{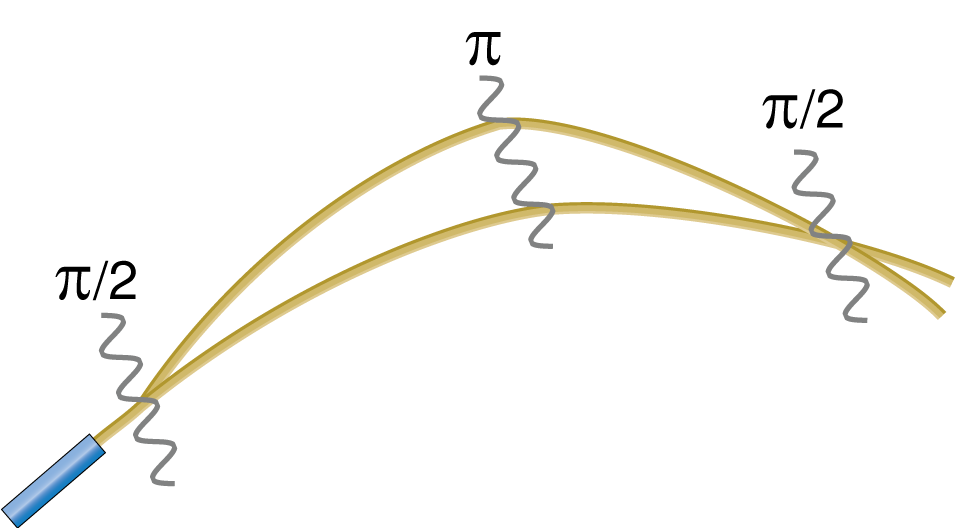}
\caption{Mach-Zehnder configuration of an atom interferometer. A first $\pi/2$ laser pulse splits the atom path in two. Subsequent $\pi$ pulses acting as atom mirrors recombine the paths that are brought to interference by a second $\pi/2$ pulse.}
\label{fig:atomzehnder}
\end{figure}

Atom interferometers contain a source of ultracold atoms that are released into free fall. During the fall, each atom interacts multiple times with a laser. In its simplest version, the laser-atom interactions force each atom to follow the two paths of a Mach-Zehnder type interferometer as shown in Figure \ref{fig:atomzehnder}. The first laser-atom interaction mimics a beam splitter for the atoms, two subsequent spatially separated interactions with each partial wave packet after time $T$ form the two mirrors of the Mach-Zehnder interferometer that recombine the two atom paths after an additional fall time $T$. A final atom-laser interaction at the point of recombination acts as another atom beam splitter. The atoms can now be counted in the two output ports of the second beam splitter. 
In its standard implementation, two counter-propagating lasers are required at each point of interaction to induce a Doppler-sensitive two-photon transition. However, single-laser interactions have been proposed recently as a promising way to mitigate some of the dominant noise contributions~\cite{YuTi2011, GrEA2013}. 

Atom interferometers (AIs) have also been considered as a new type of GW detector. 
In contrast to the laser-interferometric designs such as the torsion-bar antenna and the 
Michelson interferometer, 
AIs are generally not pure gravity strain meters, but sensitive to a multitude of field quantities 
including the homogeneous static gravity field, static gravity gradients, and fluctuations 
thereof~\cite{DiEA2008}. Another interferometer topology has been proposed that combines 
the benefits of freely falling atoms and long-baseline laser interferometry~\cite{DiEA2008b,HoEA2011}. 
In these schemes, two or more AIs interact with the same lasers. In this type of configuration, the 
AI itself no longer serves as a GW detector, but each AI constitutes a \emph{freely falling phase meter} for the lasers. Since the atoms are freely falling, these detectors are less sensitive to seismic perturbations, which is one of the major disturbances in conventional laser-interferometric detectors, requiring sophisticated vibration isolation engineering~\cite{abbott2004seismic, Accadia:2011jh}.

As reported previously~\cite{BaTh2012}, seismic noise is still relevant in laser-atom interferometers (LAIs), but it is strongly suppressed compared to seismic noise in standard laser-interferometric GW detectors. This is because any type of laser noise measured differentially between two freely falling phase meters (atom interferometers) is subject to a common-mode rejection to leading order, but does enter at order $\Omega L/c$, where $c$ is the speed of light, $L$ the distance between the two atom interferometers, and $\Omega$ is the signal frequency. Therefore, compared to conventional laser-interferometric detectors, the advantage of atom interferometers is that the common-mode rejection of seismic displacement is established optically rather than by seismic correlations between test masses. However, the results also show that laser-frequency noise needs to be further suppressed interferometrically, otherwise laser-frequency noise would pose a strong limit on the sensitivity of these detectors. Interestingly, this is ultimately a consequence of the fact that two counter-propagating lasers have to interact simultaneously with each atom. As discussed in~\cite{YuTi2011,GrEA2013}, atom GW detectors based on atom interactions with a single laser could ideally be free of laser-frequency noise (including the seismic noise) even without a laser interferometer. In the latter case, the detector could be built along a single baseline, which would be a great advantage for underground atom GW detectors since they could be constructed with a vertical baseline and vertically falling atoms. In contrast, the phase signal of each atom interferometer is first-order insensitive to the initial positions and velocities of the atoms, but constraints on the distribution of atom trajectories need to be fulfilled for example to sufficiently suppress noise associated with wavefront aberrations \cite{Ben2011}. 

\begin{figure}[h]
\includegraphics[width=0.9\columnwidth]{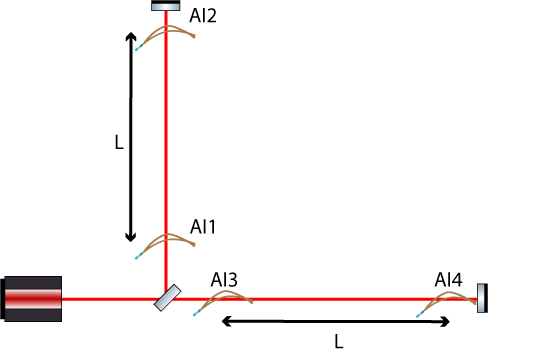}
\caption{Sketch of a possible GW detector that combines atom and laser interferometry. 
  Each of two pairs of atom interferometers (AI) measure the differential phase of the laser at a 
  distance $L$ from each other. These two differential phase signals are further subtracted from 
  each other to cancel the laser phase noise.}
\label{fig:atomlaser}
\end{figure}

The main noise contributions of atom interferometers that have been described in previous 
publications are the atom shot noise, the laser-frequency noise, Newtonian noise (see 
Section \ref{sec:Newton}), and noise associated with laser wavefront 
aberrations~\cite{Ben2011,HoEA2011b}. In the following, we will base our noise model on a 
standard LAI configuration with two perpendicular horizontal 
baselines of length 500\,m to suppress laser-frequency noise electronically similar to the 
time-delay interferometers envisioned for space-borne GW detectors such as eLISA~\cite{AET1999}. 
Seismic isolation systems are required for the main laser optics shown in Figure~\ref{fig:atomlaser} 
and for auxiliary optics forming the spatial mode filter of the input beam, but since none of the 
optics serves as test mass, the isolation requirements are less stringent.  

\begin{figure}
  \includegraphics[width=0.9\columnwidth]{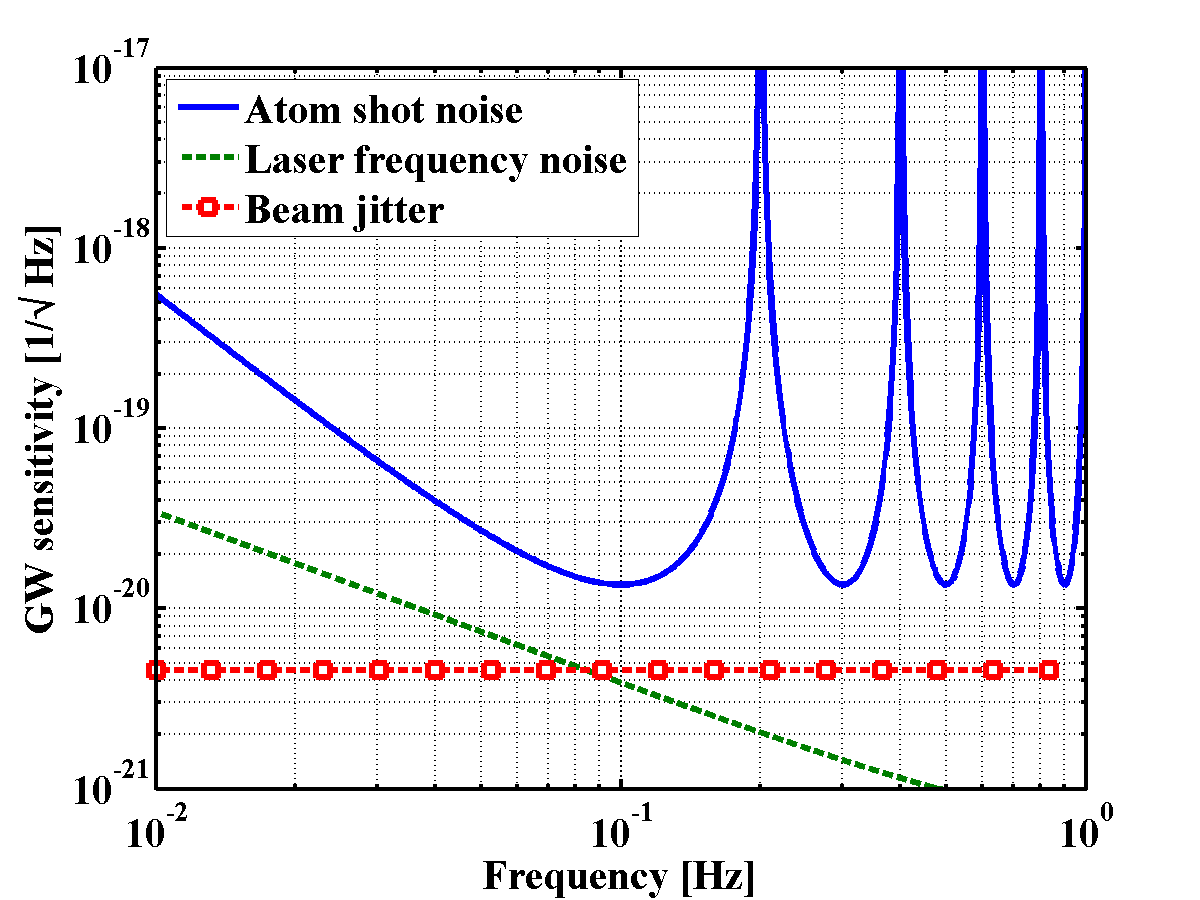}
  \caption{Sensitivity curve of the MANGO concept for a laser-atom interferometer based on the parameter values given in Table \ref{t:IFOparams}. The noise peaks are a 
    consequence of the transfer function between laser and atom phase, and are characteristic 
    for the Mach-Zehnder configuration of the atom interferometers.}
  \label{fig:atomnoise}
\end{figure}

With respect to the laser-frequency noise published in~\cite{KeEA2012}, an additional suppression of $10^5$ is assumed for the noise curve in Figure~\ref{fig:atomnoise}. Most of this suppression ($10^3$) will be achieved by performing a differential read-out between the two arms of the Michelson interferometer. However, since asymmetries between the two arms can impede noise suppression, it seems likely that MANGO sensitivity can only be achieved with an additional 100x improvement in laser frequency stabilization in the 0.1-10~Hz band relative to the level published in~\cite{KeEA2012}. This should be possible using the new generation of cryogenic laser reference cavities with crystalline mirror coatings~\cite{CoEA2013}, or building on recent progress with superradiant lasers~\cite{BoEA484}. Random displacement of the laser optics can produce excess laser-frequency noise as well as laser beam jitter that converts into atom phase noise. With respect to excess frequency noise, the requirements for seismic noise reduction around 0.1\,Hz can be about 6 orders of magnitude less stringent than they are for suspended test masses. The isolation chain up to the suspension-point interferometer (SPI) stage presented in Section~\ref{sec:Michelson} without the optical-rigid-body (ORB) and final suspension stage would provide sufficient seismic-noise suppression in the longitudinal degree of freedom (see below for additional requirements with respect to rotational degrees of freedom). The residual seismic noise $\Omega\xi/c$ ($\xi$ the optics displacement noise, $\Omega$ the signal frequency)  is less than $10^{-22}\,\rm /\sqrt{Hz}$ at $0.1\,$Hz. In this configuration the optics cannot be considered free and also the distance between laser optics is controlled over the entire detection band, and consequently optical response to GWs is suppressed. Hence, sensitivity estimates can be obtained just by considering the distance change between pairs of atom interferometers.

Static wavefront aberrations contribute to the instrumental noise if the laser beam jitters due to random tilt of the optics or the laser~\cite{Ben2011, HoEA2011b}. To provide the required alignment stability of the laser beams relative to the atoms, one first needs a stable reference, which consists of seismically isolated optics and components of the alignment control system. Then the beam jitter can be suppressed relative to the reference~\cite{Mue2005}. Also the static wavefront aberrations can be reduced by mode cleaning \cite{KwEA2012}, ultimately being limited by aberrations of the optics. Extrapolating current optics polishing and coating quality, we assume that static wavefront aberrations of $10^{-4}\,\rm rad$ should be possible. The beam jitter noise curve in Figure \ref{fig:atomnoise} was plotted with a beam jitter of $10^{-11}\,\rm rad/\sqrt{Hz}$ at 0.1\,Hz; for comparison, this is $\sim$100x better than the best angular stabilization achieved with the LIGO interferometers using differential RF wavefront sensing. Even though it seems feasible to build a control system that can suppress beam jitter down to this level, it will be very challenging to provide the seismic isolation with respect to tilt/yaw motion. This can only be achieved through passive seismic isolation or by implementing other inertial references. A solution would be to implement a multi-stage passive isolation further reducing seismic noise in all degrees of freedom. Additional suppression of beam jitter noise can be achieved by implementing adaptive optics to correct wavefront aberrations.

In addition to static wavefront aberrations that convert into atom-phase noise through beam jitter, dynamic wavefront aberrations generated by Brownian noise in the optics coatings cause additional atom phase noise. As for the beam jitter case, noise from small-scale aberrations is strongly suppressed \cite{HoEA2011b}, and we can focus on the largest scale aberration for the noise estimate, which corresponds to a spatial wavelength equal to half of the beam diameter. In this case, assuming a mirror at room temperature with a coating quality factor $Q=10^4$, the atom phase noise at 0.1\,Hz in units of GW strain is less than $\rm 10^{-24}/\sqrt{Hz}$.

Besides Newtonian noise, the most significant noise contribution is the atom shot noise governed by the flux $\eta$ of cold atoms interacting with the laser beams, and the number $n$ of photons transferred to each atom at each point of interaction with the lasers, which determines the momentum transfer from light to atoms. Since the standard-quantum-limit enforces a strong limit on the photon number in low-frequency laser-interferometric detectors, it seems feasible that atom shot noise 
can be brought to a level comparable to photon shot noise. Atom shot noise is proportional to $1/n$ and $1/\sqrt{\eta}$. The parameter values used for the noise curve in Figure~\ref{fig:atomnoise} are $\eta=10^{14}$ atoms/s (about a factor $10^6$ above current state-of-the-art~\cite{TrEA2001}) for the atom throughput, and $n=1000$ for the number of photons (about a factor of 10 above current state-of-the-art~\cite{McEA2013}). Momentum transfers with $n=102$ photons have already been realized, but without being able to measure phases~\cite{ChEA2011}.
Another option to mitigate atom shot noise is to prepare the atoms in phase-squeezed states 
through non-linear atom interactions, but atom phase-squeezing has not been demonstrated 
yet in atom interferometers. 

In summary, major technology advance and better understanding of noise sources in LAIs are 
required to achieve the sensitivity goal. Such insight can only be obtained through further 
theoretical studies, and eventually through prototyping of detectors. An important first step 
towards low-frequency GW detection would be to achieve sensitivities that would allow us to 
observe terrestrial gravity perturbations around 0.1\,Hz and to demonstrate Newtonian-noise 
subtraction at these frequencies. From Section~\ref{sec:Newton} we know that this can already 
be achieved with strain sensitivities around $10^{-16}\,/\sqrt{\rm Hz}$ (more easily in environments 
with elevated seismic and infrasound noise). This sensitivity could be achieved with a single baseline 
LAI using state-of-the-art laser-frequency stabilization~\cite{KeEA2012}. Moreover, only modest 
seismic-noise suppression by about a factor 1000 to avoid excess laser-frequency noise, and a 
modest increase of momentum transfer to $n=100$ are sufficient, while using already 
available atom flux. The length of the baseline would still have to be around 500\,m, which can 
be made smaller if either $\eta$ or $n$ are further increased.

\section{The Torsion Bar Antenna} 
\label{sec:TOBA}
\begin{figure}[h]
  \includegraphics[width=6cm]{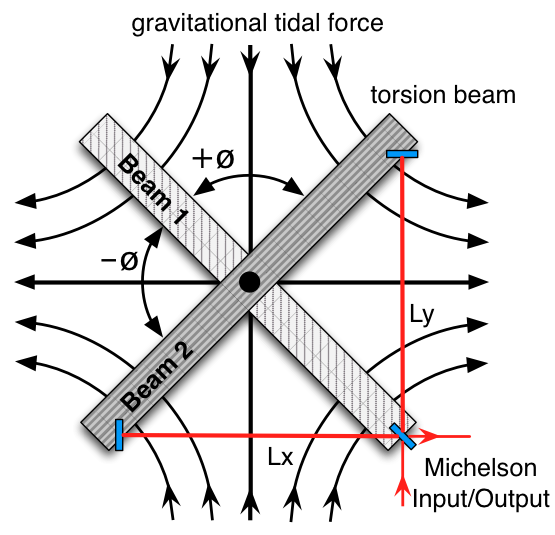}
       \caption{Interaction of TOBA's dual torsion beam configuration with GW tidal-forces.}
       \label{fig:toba_concept}
\end{figure}

A torsion-bar antenna (TOBA) is a new type of gravitational wave detector~\cite{Ando2010.PRL}. The tidal-force fluctuations caused by GWs are observed as differential rotations between two orthogonal bars, independently suspended as torsion pendulums. They share the same suspension point, have their axis of rotation co-linear and center-of-mass co-incident. This is a crucial design feature and will provide a high level common mode rejection ($\sim$1 part in 1000) from mechanical noise. Shown in Figure~\ref{fig:toba_concept}, an incoming gravitational wave, incident into the page, will rotate the beams differentially. The linear distance between the ends of the beams, Lx and Ly, will change. The differential length changes will be measured in the same way as in the long baseline gravitational wave detector (LIGO, VIRGO). Any linear pendulum motion between the beams will be registered as a common mode motion, to which the Michelson is insensitive. 


\subsection{The torsion pendulum}
\label{sec:TOBAbar}
The anticipated design for a large-scale TOBA detector has a suspended mass of $\sim10^4$\,kg (10\,m long $\times$ 0.6\,m diameter), made from a high quality low-loss material compatible with cryogenics like silicon, or Aluminium 5056. The aspect ratio of the bar is optimized to maximize the eigenfrequency of the second bending mode, to be above 10\,Hz, which generates a differential displacement between the two ends. The torsion wires need to be made from a similar high quality factor material. The fundamental torsion frequency will be around $30\,\mu$Hz. Increasing the length of the bar will improve the overall sensitivity, yet it will also increase the thermal noise associated with its internal modes. Constructing the bar with a dumbbell shape is another possibility to increase the bar's inertia by a factor of 3.
The detector will operate at cryogenic temperatures to mitigate the thermal noise (suspension thermal noise in particular). 

\begin{figure}[h]
  \includegraphics[width=\columnwidth]{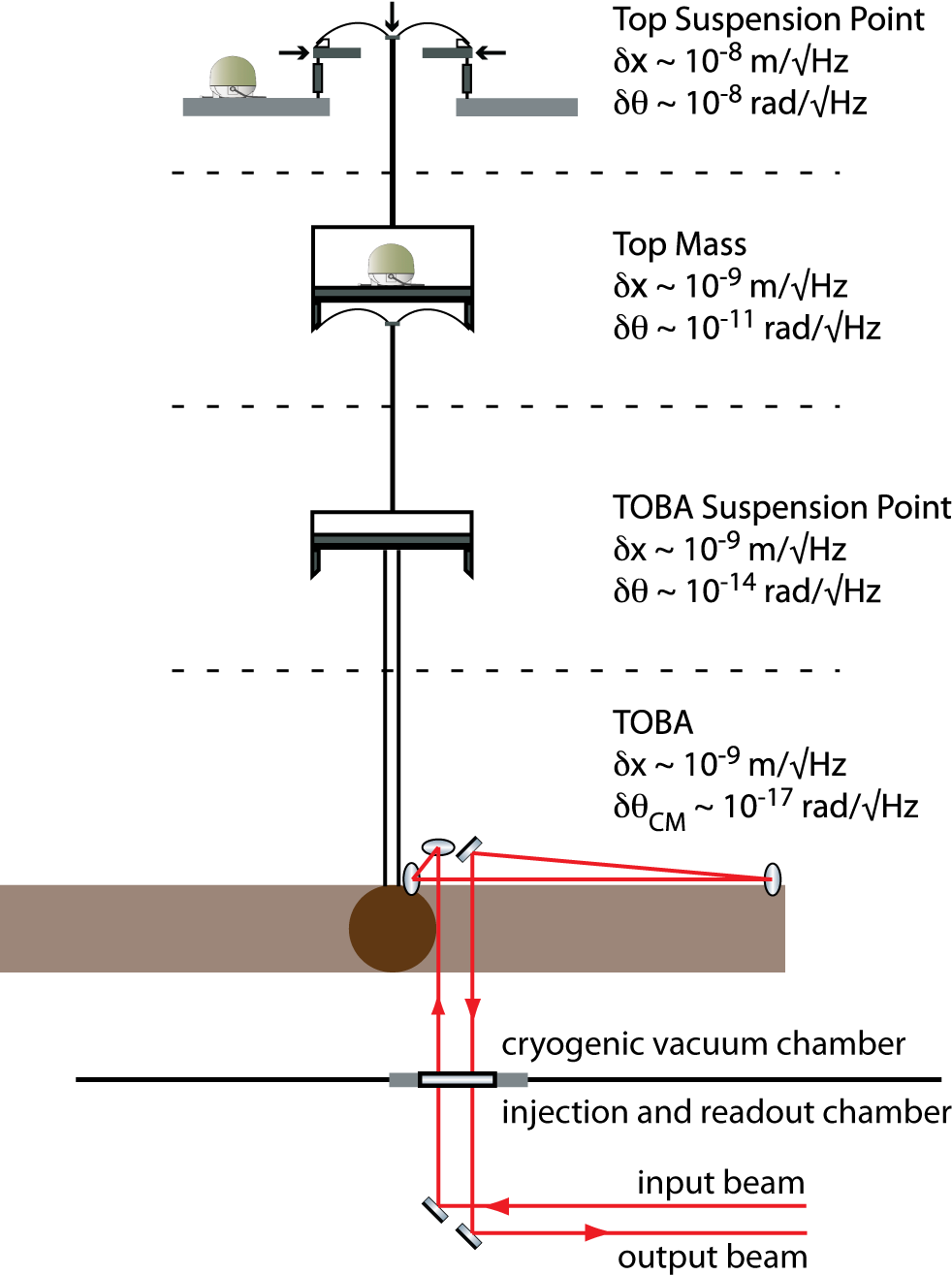}
       \caption{Schematic overview of the TOBA suspension design, with the horizontal bar at the bottom and the second bar indicated with the darker circle coming out of the page.}
       \label{fig:TobaCartoon}
\end{figure}

One of the challenges is to mount the two bars such that there is no cross coupling between the torsional and other modes. One approach is to drill a hole in the middle of one bar, while narrowing the center of the other bar. Other mechanical configurations are under investigation, such as adjusting the height of the suspension points on the bar, while maintaining the location of the center-of-mass. Figure~\ref{fig:TobaCartoon} shows a complete schematics overview of the TOBA suspension design. Here the two bars are illustrated as a solid beams.

As an alternative to solid bars, the bars can be made of a light open frame structure with large masses at the ends. This will be detrimental for the thermal noise (and low eigenmode frequencies), however linear cavities along the length of the structure can monitor the modal-displacements between the end masses.  A feedback system using inertial actuators (e.g. mass on a piezoelectric actuator) located at the anti-nodes of the first few structural modes can be used to damp the eigenmodes. Alternatively, recorded modal displacement can be used in a post-processing cancellation schemes.

The torsion bars are suspended from a common suspension point (TOBA Suspension Point  in Figure~\ref{fig:TobaCartoon}), improving the common mode rejection. The two bars have two suspension wires to accommodate the co-incidence of their axis of rotation. The wires will have a small separation at the suspension point and at the bar. The impact on the torsion frequency will be modest if the suspension wires are sufficiently long.




\subsection{Isolation chain}
\label{sec:TOBAsus}
The TOBA Suspension Point is suspended from a two stage isolation chain, inside a vacuum 
chamber to reduce seismic and acoustic coupling (see Figure~\ref{fig:TobaCartoon}). The base of the Top Suspension Point is mounted to the ground. The Top Suspension Point is isolated in four degrees-of-freedom (no roll or pitch DOF), via an inverted pendulum and a geometric anti-spring (GAS) filter \cite{Cella2005.NIMA}. It has actuators for each degree-of-freedom with respect to the ground. To reduce the force actuation it is desirable to have the eigenmodes up to 500\,mHz.

The Top Mass in Figure~\ref{fig:TobaCartoon}, is suspended from the Top Suspension Point via a single wire and is used as a reference to stabilize the residual seismic motion. A high-sensitivity broadband seismometer is mounted inside the Top Mass and registers any residual motion. The sensor data are used in a feedback control system to the actuators at the Top Suspension Point. The seismometer is housed in a pod to make it vacuum compatible, shielded from magnetic field noise and temperature stabilized to improve noise performance at low frequencies ($<$1\,Hz). The actuators will suppress the motion of the top mass down to the noise floor of the seismometer ($\approx 10^{-9}\,\rm m/\sqrt{Hz}$ \cite{RiHu2010}).

The TOBA Suspension Point is suspended from the Top Mass via a suspension wire and vertical blades, providing an additional 6 degrees of isolation. This acts as a reference for the various sensors and actuators to the individual bars.

To reduce the suspension thermal noise, the TOBA Suspension Point and below is cooled down to 4\,K. The whole suspension chain will be wrapped in heat shield to maintain the cryogenic temperatures.


\subsection{Interferometric Readout}
\label{sec:TOBAquantum}

The differential rotation between the two bars is obtained by measuring the distance fluctuation between the ends of the two bars. With respect to the first bar, the ends of the second bar will advance and retreat when the bars rotate (differentially). A Michelson interferometer, with the beam splitter on the first bar and end mirrors on the second bar, is used to measure the change in length, shown in Figure~\ref{fig:toba_concept}.

To reduce the effect of the mirror coating thermal noise on the readout, large beam sizes on mirrors is beneficial. 
The optical input and output beams for the Michelson come from underneath the bars, with the injection and readout optics, such as the recycling cavities, in a separate chamber. This greatly simplifies the optical configuration on the bars, and will separate the more complex optical readout from the mechanical system. 
\begin{figure}[t]
  \includegraphics[width=\columnwidth]{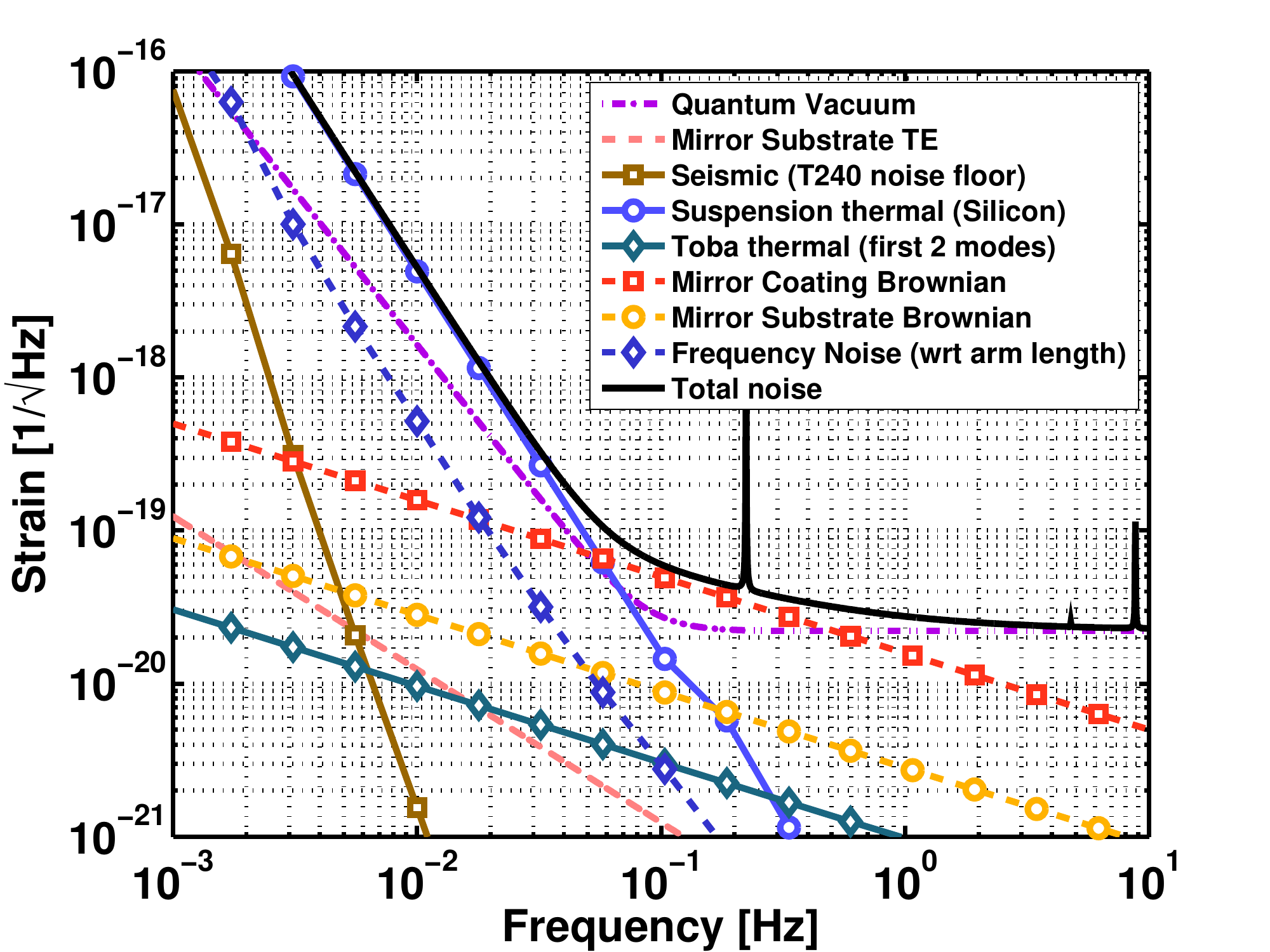}
       \caption{Noise plot of the MANGO concept for a torsion detector with a 10\,m long by 0.5\,m diameter fused silica bars (7560\,kg each) operated at 4\,K. Each bar is suspended by two 5\,m long and 2.6\,mm diameter silicon wires. The input power is set to 10\,W with no recycling cavities and a finesse of the arm cavities 313.}
       \label{fig:TobaNoiseSQZ}
\end{figure}
The final readout will be done using the DC readout technique \cite{Fricke2012.CQG} with a possible implementation of quantum-non-demolition techniques for broadband improvement of the shot noise.

Figure~\ref{fig:TobaNoiseSQZ} shows an anticipated sensitivity of a TOBA detector, operating at 4\,K. The seismic noise is set to the instrumental noise level of a broadband seismometer, $5\cdot 10^{-10}\,\rm m/\sqrt{Hz}$ at 0.1\,Hz, followed by a second pendulum stage and a 1:1000 coupling from horizontal to rotation. An estimate of the force sensitivity of the large-scale TOBA can be approximated with $\tau = I \alpha = -I \theta \omega^2 = -I (\delta x/ l)(2\pi f)^2$. With a modelled design sensitivity of $10^{-19}\,\rm m/\sqrt{Hz}$ at 1\,Hz, the torque is then $97\cdot10^3 \times (10^{-18}/10)(2\pi 10\cdot10^{-3})^2 = 4\cdot10^{-17}\,\rm Nm/\sqrt{Hz}$. At the end of the 10\,m bar a linear force as small as $\sim 10^{-17}\,\rm N/\sqrt{Hz}$ can be measured. 

\section{Michelson Interferometer} 
\label{sec:Michelson}
Another approach to low frequency ground-based gravitational-waveSom2012
detection is to modify the existing laser interferometric detector design.
These detectors are limited at low frequencies by seismic noise, thermal noise, 
and radiation pressure noise.
Though ground-based detectors all have extensive seismic
 isolation systems, seismic noise is dominant below $\sim$~5\,Hz.
As we describe below, an extension of the suspension point 
interferometer~\cite{Aso2004,Num2008} concept to many
degrees of freedom can potentially provide significant rejection of
seismic noise coupling.

To best make an optical-rigid-body (ORB) with interferometric sensing,
 a triangular configuration is chosen over the L-shape in use today.
This provides high sensitivity to all motions in the plane of the interferometer,
 making the horizontal ``stiffness'' of the ORB as high as possible.
This configuration also has other advantages as a GW detector,
 as discussed in various proposals for future detectors (ET, LISA, BBO),
 including redundancy and sensitivity to both GW polarizations.

\subsection{Pre-Isolation}
The first stages of seismic isolation for the Michelson interferometer are similar
 to those currently in use in ground-based GW detectors (e.g., Advanced LIGO).
An active pre-isolation stage reduces somewhat the noise transmitted to lower stages,
 and provides a wide range actuator for positioning the suspension chain.

A second layer of isolation is provided by low-frequency passive mechanical resonators
 (e.g., Robert's linkages for horizontal and Euler buckling springs for 
vertical~\cite{GaEA2003,WBS2002}).
These can be tuned to a few mHz to provide modest in-band isolation,
 and significant reduction of the microseism at 100\,mHz.

The target for pre-isolation is to arrive at $1 \,\rm nm/\sqrt{Hz}$ at 10\,mHz,
 and  $100 \,\rm pm/\sqrt{Hz}$ above 100\,mHz.
This motion is assumed to be present in all translational degrees-of-freedom,
 and incoherent between platforms.
\begin{table*}[ht]
    \begin{centering}
      \begin{tabular*}{0.95\textwidth}[t]{l|c r c | l | c r c}
\hline
\multicolumn{8}{l}{Michelson Interferometer}\\
\hline
Parameter & Symbol & Value & Units & Parameter & Symbol & Value & Units \\
\hline\hline
Light Wavelength & $\lambda$   & 1550   & nm      &  Substrate Young's Modulus & $Y_{\rm sub}$ &185  & GPa\\
Mirror Mass & $m$ & 600     & kg       &  Suspension Temperature & $T_{\rm sus}$    & 0.2 & K \\ 
Arm Cavity Length &  $L$ & 300     & m        & Suspension Ribbon & - & Silicon & - \\
Arm Cavity Power & $P_{\rm cav}$   & 50       & W        &  Substrate Loss Angle & $\phi_{\rm sub}$  & $3 \times 10^{-9}$ & rad \\
Beam Radius & $\omega$     & 1         & cm      &  Coating Loss Angle & $\phi_{\rm coat}$ & $2 \times 10^{-5}$ & rad \\
Detection Efficiency & $\eta$ &  0.95   & -         & Mirror Coating & - & GaAs:AlAs & - \\
Squeeze Factor & $R$ & 10       & dB       & Mirror Temperature & $T$   &   120 & K \\
\hline
\multicolumn{8}{l}{Torsion-Bar Antenna}\\
\hline
Parameter & Symbol & Value & Units & Parameter & Symbol & Value & Units \\
\hline\hline
Power & P & 10 & W     & Torsion Resonance Frequency & $\omega_{\rm tor}$ & 0.2 & mHz \\
Mirror Substrate & - & Silicon & - & Bar Length & $L_{\rm bar}$ & 10 &  m  \\
Beam Radius & $\omega$ & 2 & mm &   Bar Diameter & d & 0.5 &  m  \\
Bar Substrate & - & Fused Silica & - & Suspension Temperature & $T_{\rm sus}$ & 4 & K \\
Suspension Wire & - & Silicon & - & Bar Temperature & $T_{\rm bar}$ & 4 & K  \\
Suspension Length & $L_{\rm sus}$ & 5 & m & & & & \\
\hline
\multicolumn{8}{l}{Laser-Atom Interferometer}\\
\hline
Parameter & Symbol & Value & Units & Parameter & Symbol & Value & Units \\
\hline\hline
Arm Length & $L$ & 500 & m & Coating Loss Angle & $\phi_{\rm coat}$ & $10^{-4}$ & rad \\
Momentum Transfer & $n$ & 1000 & - & Wavefront Aberrations & $\delta\phi_{\rm wf}$ & $10^{-4}$ & rad \\
Atom Throughput & $\eta$ & $10^{14}$ & s$^{-1}$ & Beam Jitter & $\delta\alpha$ & $10^{-11}$ & rad/Hz$^{1/2}$\\
Beam Radius & $\omega$ & 1.5 & cm & & & & \\
\hline
    \end{tabular*}
    
    \caption{Interferometer parameters used for the MANGO detectors}
    \label{t:IFOparams}
  \end{centering}
  
\end{table*}

\subsection{Suspension Point Interferometer}
\label{s:MichSPI}
\begin{figure}[h]
  \includegraphics[width=\columnwidth]{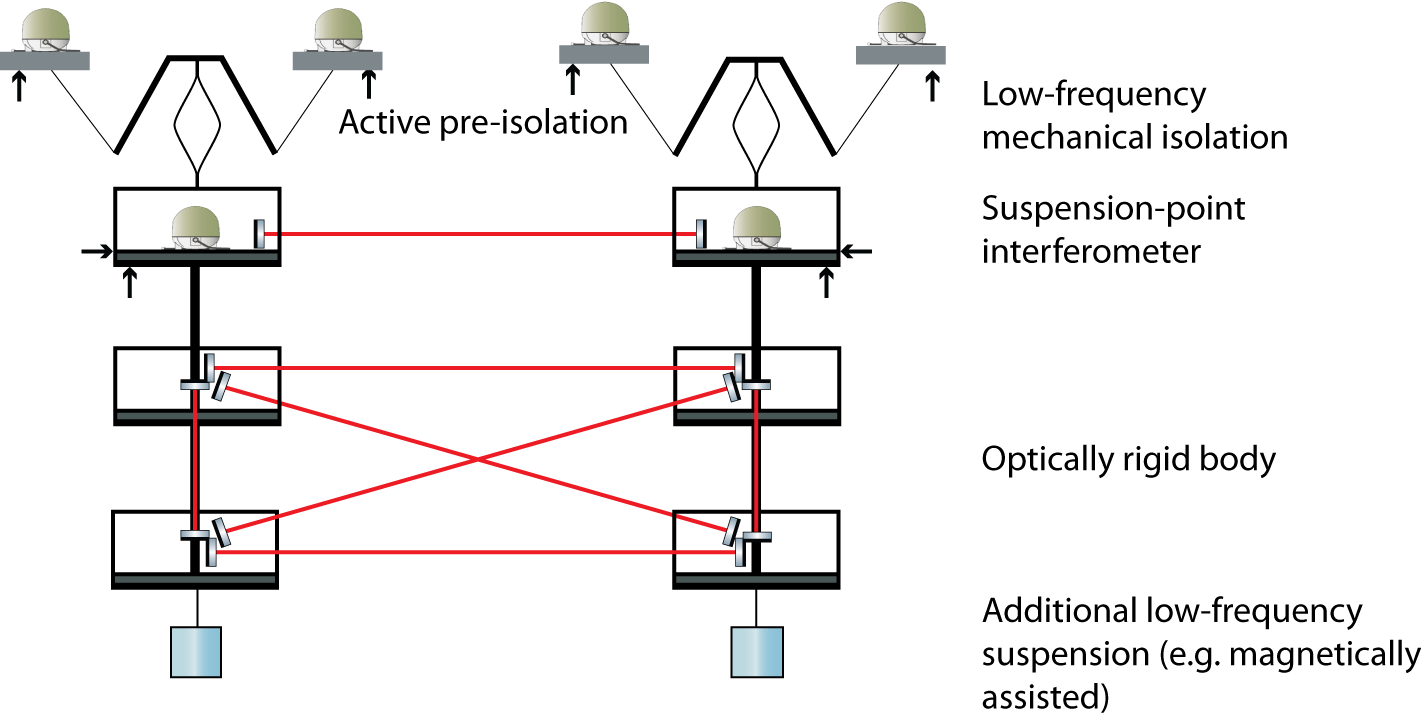}
       \caption{Seismic isolation for the Michelson low-frequency detector is provided
       by a multi-stage suspension with interferometric length and angle sensing.}
       \label{fig:MichSPI}
\end{figure}

The next layer (cf. Fig.~\ref{fig:MichSPI}) of isolation links the 3 detector platforms with Fabry-Perot cavities
 in a configuration known as a Suspension Point Interferometer (SPI).
The SPI layer serves to reduce the relative motion of the 3 platforms
 in the plane of the interferometer, and to provide interferometric alignment
 signals for the platforms.
In total, the SPI produces 3 displacement signals and 9 alignment signals,
 while the 3 platforms have a total of 18 rigid-body degrees of freedom (DOFs).
Thus, the available signals are sufficient to constrain the 3 platforms to behave as
 a single rigid-body, by removing 12 internal DOFs and leaving 6 DOFs uncontrolled
 (the SPI is clearly insensitive to translation and rotation of the 3 platforms as a rigid-body).

The alignment signals which provide the majority of the constraints,
 when coupled with small lever arms, have a sensitivity
 comparable to the displacement signals produced when the platforms are
 displaced in the plane.
That is, if we consider an SPI made of low-finesse cavities with a few 100\,mW of stored power,
 a 1\,mm lever-arm makes the $10^{-13} \,\rm rad/\sqrt{Hz}$ sensitivity
 of a wavefront sensor comparable to the
 $10^{-16} \,\rm m/\sqrt{Hz}$ shot noise limited displacement sensitivity.

The differential vertical motion (DVM) of the platforms, however, is a different matter.
DVM is detected by the SPI only through angular signals, and has an effective
 lever arm of the distance between the platforms (e.g., several hundred meters).
Designing the SPI cavities to be nearly concentric, with the radii of curvature of the mirrors
 slightly larger than half the length of the cavity,
 can increase their sensitivity
 to DVM by a factor of 10 or even 100.
This displacement noise will, however, remain
 3 to 4 orders of magnitude larger than the in-plane displacement noises,
 and only marginally lower than the noise level provided by the pre-isolators.
 
The net effect is that common motion of the 3 platforms,
 and their differential vertical motion, remain at or near the noise level
 given by the pre-isolators.
These noises will couple into later stages of the isolation chain via
 small asymmetries in the suspensions to produce motion in the plane
 of the interferometer.
A well tuned mechanical system can minimize these couplings,
 possibly to less than $10^{-4}$ with in-situ tuning, limited by thermally
 driven mechanical drifts in the suspension system.
The existence of these cross-couplings is the reason that a single layer SPI
 is not sufficient to bridge the 8-order of magnitude gap between the
 pre-isolator output noise
 and the noise level required at the test-mass suspension stage.
 
 Since greater suppression would most likely be futile, the SPI stage aims to reduce the relative motion of
  the platforms to $10^{-14} \,\rm m/\sqrt{Hz}$ at 100\,mHz, or $10^{-4}$ times the
  noise floor presented by the pre-isolator.
 If significantly better decoupling is available, the relative motion can in principal
  be further reduced to approach the
  shot noise level of the SPI around $10^{-16} \,\rm m/\sqrt{Hz}$.

\subsection{Optical Rigid Body}
\label{s:ORB}
The final layer of seismic isolation in the interferometer suspension chain
 is the optical rigid body; essentially a multi-cavity SPI which is designed to
 \emph{maximize} the coupling of vertical displacement to the readout.
The collection of resonant optical cavities which constitute the ORB
 are arranged such that any differential displacement of the platforms
 appears as a longitudinal displacement of at least one cavity.
When all of these cavities are held at their resonance points with active
 control loops, the 3 independently suspended platforms are forced to move
 as a rigid body.
Furthermore, since the ORB cavities span mechanically separated
 layers in the suspension chain, the bottom layer can be used as
 a proof-mass in an ``interferometric seismometer'',
 thereby allowing for the reduction of common motion of the ORB.
 
The aim of the ORB is to reduce the common displacement,
 and differential vertical motion, to $10^{-13} \,\rm m/\sqrt{Hz}$ at 100\,mHz
 at the bottom layer of the suspension chain.
The in-plane differential motion can then be reduced to the shot noise level of
 the bottom layer ORB cavities around $10^{-16} \,\rm m/\sqrt{Hz}$.
These are the noise levels presented as inputs to the final suspension
 stage which holds the interferometer test-masses.

It is worth noting that the ORB is in principle sensitive to gravitational waves,
 since it is made of optical cavities identical to the ones used in the test-mass
 stage interferometer.
Since the ORB control loops suppress any detected motion of the suspended
 platforms, they will also suppress any GW signal which appears within its control
bandwidth.
Thus, below the resonance frequency of the final-test mass suspension stage,
the interferometer which hangs from the ORB will be insensitive to GWs.
Above that frequency, however, the test masses are free to move relative to the ORB
and the GW signal is not suppressed.
Since the resonance of the test-mass suspension is necessarily below the GW
detection band, suppression of the GW signal by the ORB is not a problem.

\subsection{Magneto-Mechanical Suspension}
\label{sec:MagLev}
The final suspension stage of this low frequency Michelson interferometer presents
several technical challenges.
The most obvious of these is to attain a pendular resonance frequency below the band of
interest for GW detection; in our case this is 10\,mHz.

The second major challenge of the final suspension stage is to provide low thermal noise.
The thermal noise of a simple pendulum suspension, above the mechanical resonance, is given by
\begin{equation}
x_{\rm thermal} = \frac{1}{m \omega^2}\sqrt{\frac{4 k_{\rm B} T \; k \phi}{\omega}}
 = \sqrt{\frac{4 k_{\rm B} T \; \omega_0^2}{m \; Q \; \omega^5}}
\end{equation}
 where $k_{\rm B} T$ is the Boltzman constant times the suspension temperature,
 $m$ the suspended mass,  $k$ the effective spring constant, 
 $Q = 1/\phi$ the quality factor and $\phi$ the loss-angle of the restoring spring,
 $\omega_0 = \sqrt{k / m}$ the resonant frequency, and $\omega$ the measurement frequency.
To put in some rough numbers, a 1\,mHz suspension with $Q = 10^8$ holding a
 100\,kg test-mass would result in $\sim 3 \times 10^{-17} \,\rm m/\sqrt{Hz}$ thermal noise
 at 100\,mHz; this is not enough to reach the MANGO goal.

In a magnetic or magnetically assisted 
suspension~\cite{Monica:Magnetic, Drever:Magnetic, Giles:SuperTorsion}, the thermal noise may
 not come only from the restoring force of the suspension, but also
 from the deformation of the suspension element which counters the force of gravity.
Magnetic suspensions may also have losses due to magnetostriction in the support
 magnets, and eddy current damping in conductive suspension components. It may
be possible to avoid some of these issues by using an electrostatic suspension 
instead~\cite{Giaz:1998}.

The third major challenge is matching; the common motion of the ORB can become
 differential motion of the test-masses if the restoring forces of the test-mass suspensions
 are not perfectly matched.
Numerically speaking, the suspensions must be matched
 well enough to reject the $10^{-13} \,\rm m/\sqrt{Hz}$ common motion of the ORB at a
 level of $\sim 10^{5}$ to prevent it from spoiling the detector sensitivity.

\subsection{Detector Sensitivity}
\label{sec:MichSens}
In order to reach a strain noise level of less than $10^{-20}\,\rm/\sqrt{Hz}$ at 0.1\,Hz, major developments in suspension and quantum-noise technology are required. The parameter values for the Michelson MANGO configuration are summarized in Table \ref{t:IFOparams}. 
\begin{figure}[t]
  \includegraphics[width=\columnwidth]{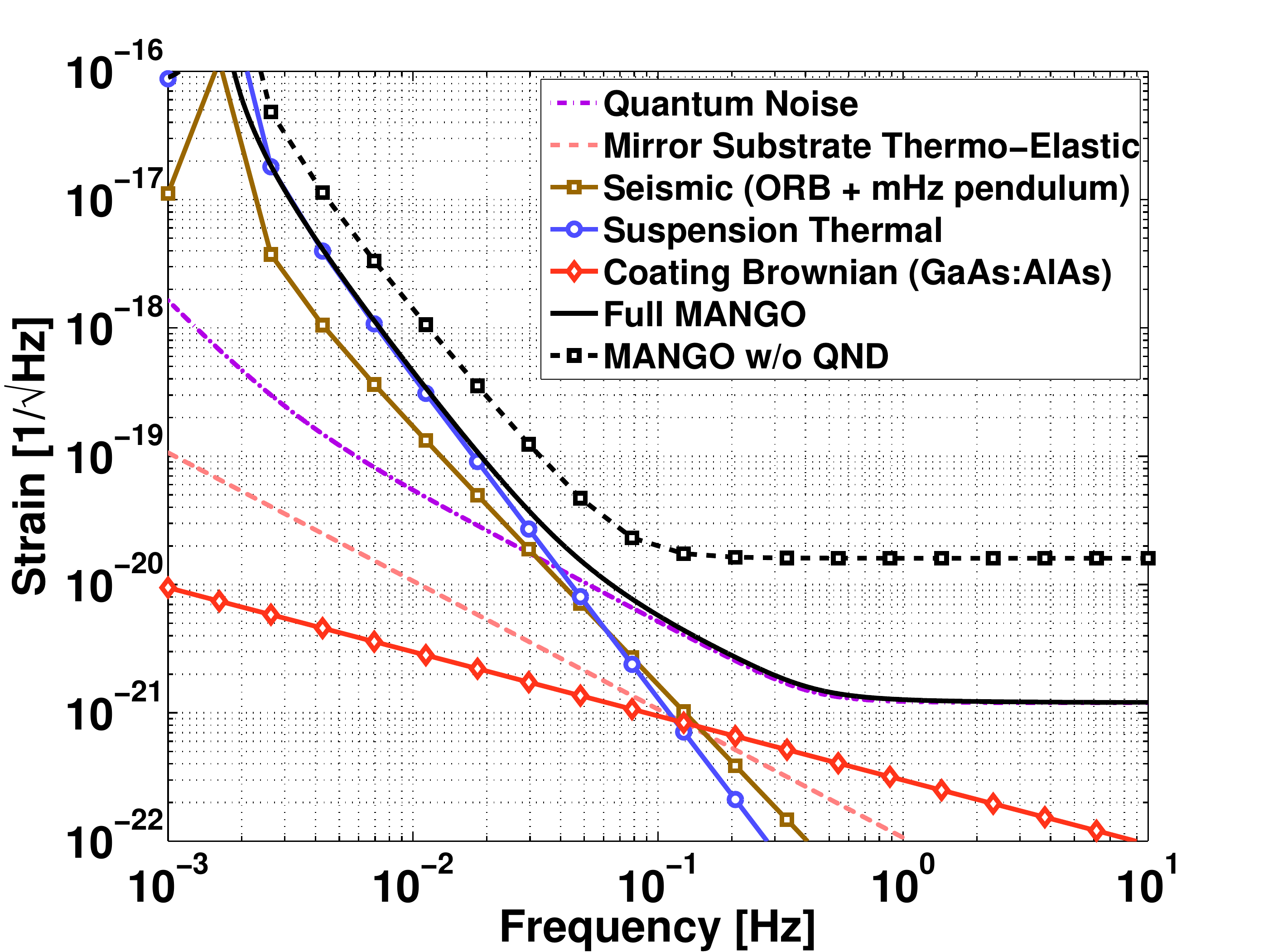}
       \caption{Strain noise of the MANGO concept for the dual-recycled, Fabry-Perot Michelson 
         interferometer with 300\,m arms, and a speed-meter cavity used to suppress the radiation 
         pressure noise. All fundamental noise contributors are included. This noise budget is 
         intended to represent the best sensitivity currently imaginable. For comparison, the 
         sensitivity is also shown for a less ambitious design with respect to radiation-pressure 
         noise reduction.}
       \label{fig:MichNoise}
\end{figure} 
The quantum noise is achieved by applying quantum-non-demolition (QND) 
techniques such as a speed-meter design. However, these can only be realized by means 
of extreme low-loss, small-bandwidth (i.e.~$\sim$0.1\,Hz) optical resonators, or alternatively, 
phenomena in light-atom interactions such as 
electromagnetically induced transparency (EIT) \cite{Mik2006, EIT:RMP2005}
could potentially fulfill the same purpose. A less ambitious detector design would not rely on 
QND techniques. Reducing the mirror mass to 90\,kg, and optimizing the laser power, a strain 
sensitivity can be achieved that is about a factor 10 smaller above 0.1\,Hz 
and is shown in Fig.~\ref{fig:MichNoise} as ``MANGO w/o QND''.

\section{Newtonian Noise}
\label{sec:Newton}

One of the foremost problems of ground-based GW detectors operating at frequencies below 10\,Hz and common to all detector types is the Newtonian noise (NN). Newtonian noise is generated by a fluctuating terrestrial gravity field. In the following, we will discuss some of the known contributions to NN, and conclude with a brief review of coherent NN subtraction. 

It should be noted that atom interferometers can have additional Newtonian noise terms compared to laser-interferometric GW detectors \cite{DiEA2008,VeVi2013} since the phase evolution of matter waves depends on the gravity potential. These terms add to the NN response from distance changes between two atom interferometers. However, at least for the LAI configuration discussed in Section \ref{sec:atom} based on Mach-Zehnder atom interferometers, one finds that the total NN in units of GW strain is identical to NN in laser-interferometric GW detectors (if this was not so, then NN could be coherently subtracted from a LAI using data from a collocated, equally long laser-interferometric GW detector, and vice versa).

\subsection{Seismic and Atmospheric NN}
The two main contributions to NN are produced by the ambient seismic field \cite{Sau1984,HuTh1998,BeEA1998}, and density fluctuations in the atmosphere \cite{Cre2008}. Even though the main focus of these publications is to provide NN estimates for the LIGO and Virgo detectors, it is possible to extend the models to lower frequencies. 

To obtain an accurate model of seismic NN, one needs detailed information about the seismic field. In some sense, seismic NN estimation below 1\,Hz is easier since the properties of the seismic field do not depend significantly on detector depth, and also the seismic field can often be understood by studying data from large-scale seismic networks operated by seismologists without the necessity to carry out additional site studies. Whereas body waves can dominate the seismic field at higher frequencies especially at underground or remote sites, the dominant contribution below a few tens of a Hz is almost always the Rayleigh-wave field, which is consistent with our understanding of seismic sources being mostly located at the surface (or at shallow depths relative to the length of Rayleigh waves) \cite{BoEA2006}. Therefore, the NN estimate presented here will be derived from the Rayleigh-wave field. The equation for the gravity perturbation of a single test mass at height $h$ above ground by a plane Rayleigh wave is given by:
\beq
x(\Omega) = -2\pi\irm \cos(\theta)G\rho_0\gamma_{\rm R} \xi_z(\Omega)/\Omega^2\exp(-\Omega h/c_{\rm R}),
\label{eq:dispnnray}
\eeq
where $\xi_z(\Omega)$ is the vertical displacement amplitude measured at the surface directly beneath the test mass, $\gamma_{\rm R}\approx 0.83$ is a material dependent factor that accounts for the partial cancellation of NN from surface displacement due to the sub-surface compressional wave content of the Rayleigh-wave field, $c_{\rm R}\approx 3.5\,\rm km/s$ is the speed of Rayleigh waves, $\theta$ is the angle between the horizontal direction $x$ along which the test-mass displacement is calculated and the direction of propagation of the Rayleigh wave, $\rho_0$ is the mean mass density of the ground, and $G$ is Newton's gravitational constant. At low frequencies the exponential term is approximately equal to 1 and the gravity perturbation does not depend explicitly on the Rayleigh-wave speed anymore. In this case the same equation can also be used as an approximation for underground gravity perturbations. The exact expression for underground gravity perturbations from Rayleigh waves as should be used for underground detectors at shallow depth operating at higher frequencies is more complex and involves details about the geometry of the cavity that hosts the detector. An expression similar to equation (\ref{eq:dispnnray}) is obtained for gravity perturbations along the vertical direction (without the $\pi/2$ phase shift). Finally, the factor $\gamma_{\rm R}$ would have a different value for Rayleigh overtones \cite{HuTh1998}. Here we will assume that the dominant waves are fundamental Rayleigh waves. 

In contrast to the advanced detectors that will sense gravity perturbations as differential displacement noise that is uncorrelated between the test masses, terrestrial low-frequency detectors will sense gravity gradients since the length $L$ of the detector arms is much smaller than the length of a seismic wave. Therefore, $\Omega L/c_{\rm R}\ll 1$, and the gravity gradient perturbation along the horizontal direction $x$ is obtained by multiplying equation (\ref{eq:dispnnray}) with $\cos(\theta)\irm\Omega L/c_{\rm R}$. It follows that the Newtonian strain noise $x/L$ (legitimately deserving the name gravity-gradient noise at low frequencies) is independent of the arm length. 

Creighton \cite{Cre2008} describes several types of atmospheric NN. In this paper we will focus on gravity perturbations produced by infrasound waves. It is not obvious that infrasound NN is the dominant contribution since there are no accurate models for most atmospheric gravity perturbations at low frequencies. However, extending the Creighton models naively to lower frequencies and assuming that the detectors are located sufficient far underground, other contributions to the atmospheric NN become insignificant since their noise spectral densities fall rapidly with increasing distance to the test masses. Infrasound waves are the analog of compressional seismic body waves propagating in media with vanishing shear modulus. As for the seismic NN, we first calculate the gravity perturbation from a single plane infrasound wave. The density perturbation of an infrasound wave can be written as:
\beq
\delta\rho = \frac{\rho_0}{\gamma}\frac{\delta p}{p_0}
\eeq
where $\gamma$ is the adiabatic coefficient of air, and $\rho_0$ is the mean air density. The relative pressure fluctuations $\delta p/p_0$ can be taken from published measurements \cite{KiLe1970}. The infrasound wave is incident on Earth's surface at an angle $\theta$ with respect to the normal of the surface and reflected from it without energy loss. Then the horizontal gravity perturbation at a depth $z_0$ reads
\beq
x(\Omega) = -4\pi\irm\sin(\theta)\cos(\phi)G\,\delta\rho\, c_{\rm IS}\exp(\sin(\theta)\Omega z_0/c_{\rm IS})/\Omega^3
\eeq
with $z_0\leq 0$, $c_{\rm IS}$ is the speed of the infrasound wave, and $\phi$ is the angle between the horizontal component of the propagation direction and the direction of test-mass displacement $x$. As for seismic NN, the low-frequency infrasound strain noise is independent of the arm length of the detector. 

The reduction of infrasound NN with depth depends on the angle of incidence. Similar to the case of Rayleigh waves, it is the apparent horizontal wavelength that determines the exponential reduction. Infrasound waves that propagate nearly horizontally produce gravity perturbations that have a large projection onto the horizontal direction $x$, but the gravity perturbation falls rapidly with depth. Gravity perturbations from infrasound waves that travel almost vertically cannot be efficiently reduced by going underground, but they also have a very small projection onto the direction $x$. This feature needs to be investigated more carefully in the future since it is well known that the infrasound field is highly anisotropic at lower frequencies \cite{HeEA2012}. However, it should be clear that for realizable detector depths the exponential reduction will not be very significant in general. Before we present the noise curves for the seismic and infrasound NN, we summarize the underlying simplifications:

\emph{Seismic Newtonian noise}
\begin{itemize}
\item {\it Integration is carried out over the seismic field in a half space.} Newtonian noise at the lowest frequencies may depend on Earth's curvature.
\item {\it The field is dominated by fundamental Rayleigh waves.} Especially with respect to NN mitigation, one needs to consider possible contributions from body waves and Rayleigh overtones.
\item {\it Effects of underground cavites on NN are neglected.} Underground detectors in cavities may also be sensitive to gravity perturbations from shear waves \cite{HaEA2009b}. Seismic NN in underground detectors depends on the geometry of the cavity, and scattered waves contribute to NN. The latter two effects should be negligible at low frequencies. 
\item {\it Rayleigh waves have frequency-independent speed.} In reality, Rayleigh waves can show strong dispersion \cite{BoEA2002ch2} also below 1\,Hz. The speed of continental Rayleigh waves lies within 2\,km/s -- 4\,km/s between 10\,mHz and 1\,Hz. However, since seismic NN at low frequencies does not depend significantly on the speed of seismic waves, implementing a realistic dispersion should not alter the results very much.
\item {\it Propagation-direction averaged NN is calculated assuming an isotropic seismic field.} It is well known that the seismic field can show significant anisotropies especially at low frequencies \cite{Ces1994}.
\end{itemize}

\emph{Infrasound Newtonian noise}
\begin{itemize}
\item {\it Integration is carried out over the infrasound field in a half space.} The thickness of the atmosphere can be a fraction of the length of infrasound waves. For this reason it should be expected that infrasound NN is significantly smaller below 0.1\,Hz than reported in this paper. In addition, infrasound waves are reflected from layers of the atmosphere (i.~e.~the stratosphere or thermosphere) at characteristic angles \cite{HeEA2012}. Newtonian noise at lowest frequencies may depend on Earth's curvature.
\item {\it Mean air density, air pressure, and speed of infrasound waves do not change with altitude.} 
\item {\it The speed of sound is frequency independent.} There are no studies of the dispersion of atmospheric infrasound at low frequencies (especially as a function of altitude). For a given infrasound field, dispersion has a weak effect on NN below 1\,Hz. 
\item {\it The atmosphere does not move.} Winds play an important role in the propagation of infrasound leading to characteristic patterns in the field \cite{HeEA2012}. It is unclear if wind in relation to infrasound waves has additional consequences for NN apart from the fact that wind can be a local source of infrasound when interacting with surface structure.
\item {\it Propagation-direction averaged NN is calculated for an isotropic infrasound field.} Isotropy is certainly an unrealistic assumption as mentioned before.
\end{itemize}

\begin{figure}[t]
\centerline{\includegraphics[width=\columnwidth]{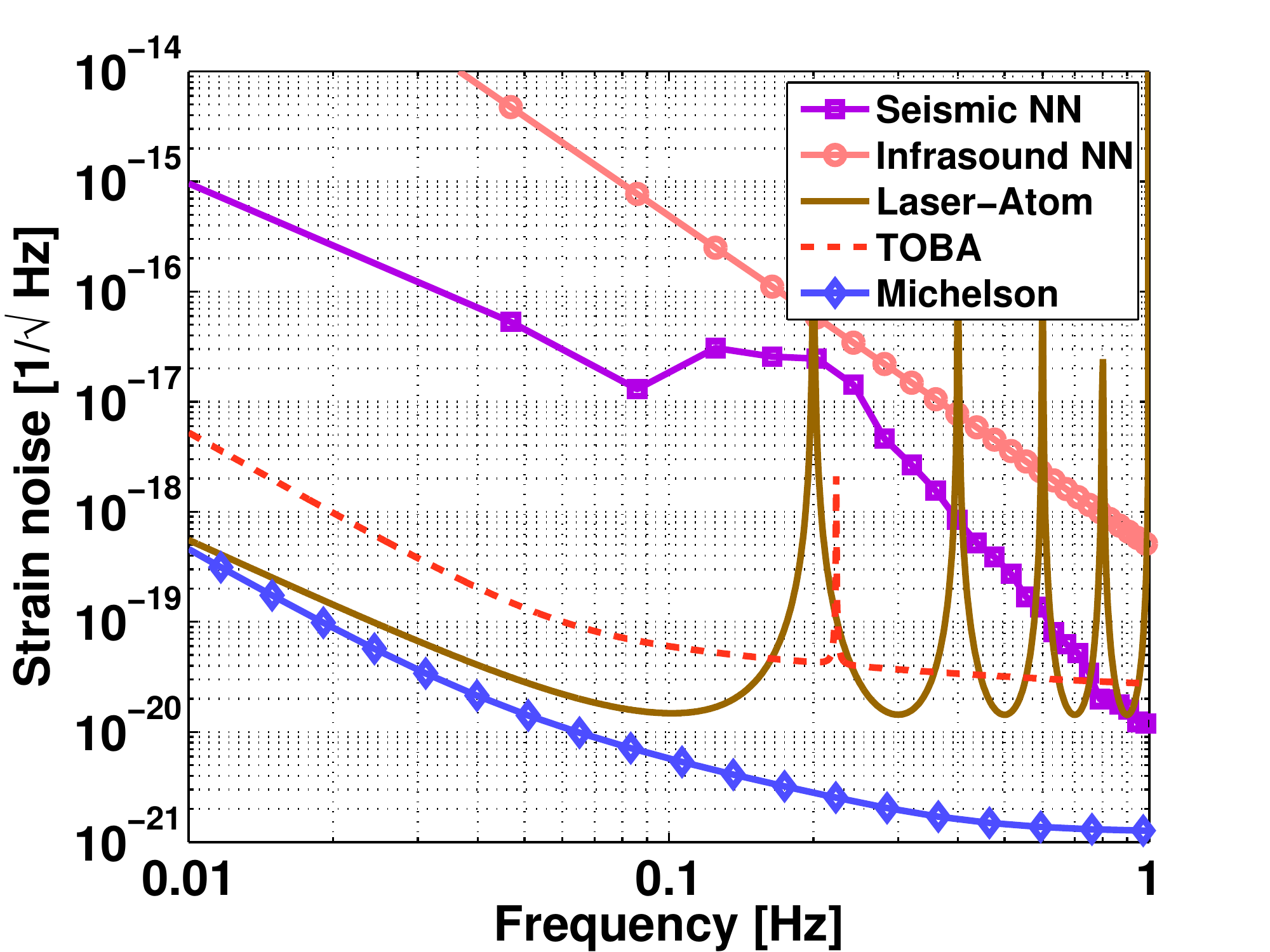}}
\caption{Seismic Rayleigh wave and atmospheric infrasound NN together with the sensitivity curves of the three MANGO concepts.}
\label{fig:Newtonian}
\end{figure}

Using the seismic spectrum published in \cite{HaEA2010} and a fit to the pressure spectrum published in \cite{KiLe1970}, we obtain the NN curves presented in Figure \ref{fig:Newtonian}.
As a final remark we want to point out that both seismic and infrasound NN have lower limits since seismic and infrasound spectra both lie above global low-noise models \cite{Pet1993,BBB2005}. Therefore, in terms of site selection, the goal should be to identify a site where both spectra are close to the respective low-noise models.

\subsection{Gravity transients}

The GW community has not paid much attention to terrestrial gravity transients in the past except for a paper on anthropogenic noise focussing on surface detectors such as LIGO or Virgo \cite{ThWi1999}. The reason for this is that gravity transients can be eliminated in high-frequency detectors simply by avoiding abrupt changes in velocity of moving objects and humans within a zone of about 10\,m radius around the test masses. The situation is very different for low-frequency detectors. Even though the terrestrial transient landscape is completely unknown and difficult to model in many cases, it is possible to identify potentially significant contributions. 

\paragraph{Newtonian noise from uniformly moving objects}
We consider the case of an object that is moving at constant speed $v$ along a straight line that has distance $r_j$ to a test mass at closest approach. Therefore, the vector $\vec r_j$ pointing from the test mass to the closest point of approach is perpendicular to the velocity $\vec v$. The closest approach occurs at time $t_j$. As before, we express the result in terms of the Fourier amplitude $x_j(\Omega)$ of test-mass displacement:
\beq
\begin{split}
&x_j(\Omega)=\\
&\quad\frac{2Gm}{v^2\Omega}\Big(K_1(r_j\Omega/v)\cos(\alpha)+\irm K_0(r_j\Omega/v)\cos(\beta)\Big)\e^{\irm\Omega t_j}
\end{split}
\eeq
Here, $m$ is the mass of the moving object, $\alpha$ is the angle between $\vec r_j$ and the arm, $\beta$ is the angle between $\vec v$ and the arm, and $K_n(x)$ is the modified Bessel function of the second kind. In all relevant cases, the argument $x=r_j\Omega/v$ obeys $x\gg|n^2-1/4|$ so that the modified Bessel functions can be expanded according to:
\beq
K_n(x)\approx \sqrt{\frac{\pi}{2x}}\e^{-x}\left(1+\frac{4n^2-1}{8x}+\ldots\right)
\label{eq:approxK}
\eeq
The moving object could be a car, a person, or a quasi-static density fluctuation in the atmosphere localized within a cell and transported by wind. In this last case, one would consider a spatial distribution of many cells with typical quasi-static density perturbation and volume determined by a spatial correlation function \cite{Cre2008}. 

In fact, one motivation to build low-frequency detectors underground comes from this type of gravity perturbation. Evaluating a few examples, one finds that the associated NN would completely dominate the signal if it was not for the exponential suppression in equation (\ref{eq:approxK}), which is effective especially for underground detectors. The threshold frequency $f_0$ above which NN from uniformly moving objects can be neglected is given by $f_0\approx v/(2\pi r_j)$. At the surface, one could imagine constructing an environmental shield around test masses with radius of about 10\,m, so that typical threshold frequencies are close to 0.1\,Hz almost independent of the object's mass $m$. So even an animal running at a straight line past the buildings of a surface detector could potentially generate significant NN up to the threshold frequency. Therefore, the only feasible solution to this problem is to build the detector several hundred meters underground and push $f_0$ below the detection band for all conceivable speeds $v$. As NN from an uncontrolled environment is avoided by increasing the distance between objects and test masses, NN control can in principle be achieved by enforcing a strict speed limit of all objects near test masses.

\paragraph{Newtonian noise from oscillating objects}
Isolated oscillating objects cannot exist as there must always be a reaction force on another object. For example, a shaking tree will transfer momentum to the ground generating seismic waves that are correlated with the motion of the tree. Therefore, the full problem of gravity perturbations from oscillating objects is difficult to analyze. The aim of this section is to provide NN estimates from the oscillating object itself without including reaction terms.

We will calculate the strain of the perturbation measured by two test masses at distance $L$ to each other forming an interferometer arm in the direction of the unit vector $\vec n$. A point mass $m$ is assumed to oscillate with amplitude $\vec\xi$ much smaller than its distance to the test masses. So we will always linearize the equations with respect to $\xi$. Then the displacement of the first test mass has the well known dipole form
\beq
x_1=-\frac{Gm}{r_0^3\,\Omega^2}\,\big((\vec\xi\cdot\vec n)-3(\vec e_r\cdot\vec n)(\vec\xi\cdot\vec e_r)\big)
\eeq
where $\vec e_r$ is the unit vector pointing from the first test mass to the object, and $r_0$ is the mean distance between them. The acceleration of the second test mass expressed in terms of the same unit vector $\vec e_r$ reads
\beq
\begin{split}
x_2&=-\frac{Gm}{r_0^3\,\Omega^2}\frac{1}{(1-2\lambda(\vec e_r\cdot\vec n)+\lambda^2)^{5/2}}\\
&\qquad\cdot\big((\vec\xi\cdot\vec n)-3(\vec e_r\cdot\vec n)(\vec\xi\cdot\vec e_r)\\
&\qquad\quad+\lambda(3(\vec\xi\cdot\vec e_r)+(\vec e_r\cdot\vec n)(\vec\xi\cdot\vec n))-2\lambda^2(\vec\xi\cdot\vec n)\big)
\end{split}
\eeq
with $\lambda\equiv L/r_0$. We evaluate the strain for the case of an oscillating object at the surface directly above the first test mass that is located at a depth $r_0$. In this case $\vec e_r\cdot\vec n\approx 0$ and the strain simplifies to
\beq
\begin{split}
h &=(x_2-x_1)/L\\
&=-\frac{Gm}{r_0^3\,\Omega^2 L}
\left(\frac{(1-2\lambda^2)(\vec\xi\cdot\vec n)+3\lambda(\vec\xi\cdot\vec e_r)}{(1+\lambda^2)^{5/2}}-\vec\xi\cdot\vec n\right)
\end{split}
\label{eq:oscstrain}
\eeq
One can see that for small detectors with $\lambda\ll 1$, the strain in equation (\ref{eq:oscstrain}) is proportional to $\lambda$ for vertical oscillations, or $\lambda^2$ for horizontal oscillations, which makes the strain disturbance independent of the distance $L$ between the test masses or proportional to $L$. In the latter case we have the uncommon situation that strain noise increases with detector length. 

We conclude this section with an estimate of NN from the sway of a single tree \cite{FlWi1999a,WiFl1999,FlWi1999b}. We assume that the tree crown displacement can be approximated as horizontal and that the test masses are located 1\,km underground forming an interferometer arm of 20\,m length. The natural frequency of a $h=15$\,m tall tree is about 0.4\,Hz. We assume the stem diameter at breast height to be $\rm dbh=0.3\,m$ so that the parabolic estimate of its mass is about $m=\rho\pi/2({\rm dbh}/2)^2h\approx 450\,$kg with a density $\rho=850\,\rm kg/m^3$. Then the strain disturbance as time-domain amplitude is given by
\beq
h=\frac{9G(m/2)}{2r_0^5\,\Omega^2}(\vec\xi\cdot\vec n)L\approx 10^{-22}
\label{eq:tree}
\eeq
assuming that effectively only half of the tree mass is displaced and that the displacement amplitude at the natural frequency is 0.5\,m in the direction of the arm. This strain value seems sufficiently small, but gravity perturbations from multiple trees could potentially add coherently. 

\paragraph{Newtonian noise from fault rupture}
Teleseismic events can cause an immediate gravity perturbation in low-frequency GW detectors in addition to a delayed perturbation from seismic waves generated by these events that pass the detector. Between 10\,mHz and 1\,Hz, earthquakes and major explosions such as the eruption of volcanoes are examples of sources of strong gravity perturbations. Here we are interested in fault ruptures since the rate of events with significant magnitude can be very high in certain regions. 

The prompt gravity perturbation from fault rupture can in principle include several distinct transients associated with lasting density changes near the fault that are built up during the fault rupture, compressional waves generated by the event, and also contributions from strong surface displacement at the epicenter depending on the event depth. The relative strength of these transients depends significantly on the location and orientation of the GW detector with respect to the fault plane. Details will be presented in an upcoming publication. Here we will focus on a simple estimate of the fault-rupture detection horizon of MANGO based on the well known lasting gravity change produced by earthquakes \cite{Oku1992,Oku1993} that has been observed in multiple occasions \cite{ImEA2004,CaSa2013}.

The measured gravity strain depends on the location of the detector with respect to the fault and slip orientation. For a strike-slip event at 1000\,km distance, with fault length and width equal to 12\,km, the center of the fault at 25\,km depth, slip size of 1\,m, and ideal detector location, we obtain a lasting change in radial gravity strain of $4\times 10^{-18}$.  Rupturing a fault of this size would take about 2\,s, which corresponds approximately to a magnitude $M=6$ earthquake. As the event corner frequency would be about 0.5\,Hz, this perturbation could easily be seen in the data. Similar results are obtained for dip-slip events and arbitrary fault orientations.

\subsection{Newtonian noise subtraction}
Since terrestrial gravity perturbations cannot be fully avoided, alternative noise-mitigation strategies need to be developed. One idea is to monitor the density fluctuations around the test masses. Most importantly, this means to measure seismic waves and atmospheric infrasound by means of sensor arrays. The sensor data can then be used to produce a coherent subtraction filter for NN (i.~e.~a Wiener or adaptive filter \cite{Cel2000,DHA2012}). This technique seems to be very attractive since clearly a high number of sensors like seismometers and infrasound microphones should make it possible to subtract a large part of the NN. However, as we will demonstrate in this section, it is uncertain whether sufficiently sensitive seismic and infrasound sensors can be provided. 

Even though infrasound waves are the atmospheric analog of compressional body waves, it is not possible to achieve high infrasound NN subtraction with a single microphone as suggested for compressional waves in \cite{HaEA2009b} using a single seismic strainmeter. The main reason is that microphones respond to infrasound waves independently of the direction of propagation, whereas seismometers measure ground displacement in certain directions. In addition, it is generally impossible to achieve significant broadband subtraction of Rayleigh seismic NN with a single seismic sensor, independent of the type of seismic sensor that is used \cite{DHA2012}.

\begin{figure}[t]
\includegraphics[width=\columnwidth]{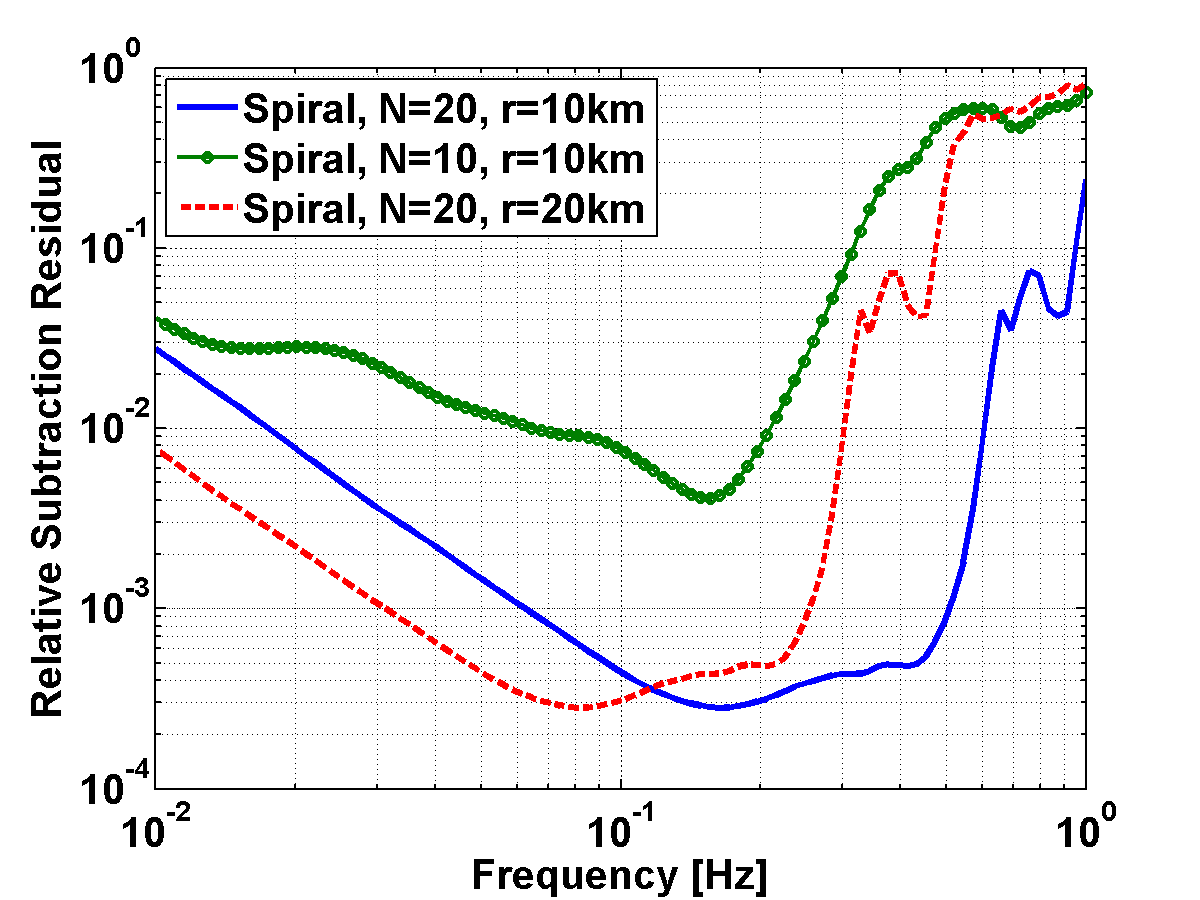}
\caption{Residuals of Rayleigh-wave gradient NN subtraction for double-wound spiral arrays using seismometers with $\rm SNR=1000$. Results are presented for different numbers $N$ of seismometers, and different array radii $r$.}
\label{fig:resRayleigh}
\end{figure}

Figure \ref{fig:Newtonian} shows that seismic and atmospheric NN would have to be reduced by large factors to achieve sensitivity goals with respect to NN. Performance of NN subtraction over a band of frequencies not only depends on the sensitivity of the auxiliary sensors, but also on the design of the sensor array. Here we will present results for the atmospheric and Rayleigh seismic NN subtraction using Wiener filters as outlined in \cite{BeEA2011}. In Figure \ref{fig:resRayleigh}, the three curves represent relative subtraction residuals for three spiral arrays.
The calculation is based on an isotropic field of Rayleigh waves. The sensors measure vertical ground displacement with $\rm SNR=1000$. The detector length is $L=200\,$m. Array density determines the highest frequency up to which NN can be subtracted. At low frequencies, subtraction performance declines, because a larger fraction of the seismic signal leads to common-mode gravity perturbations that are rejected by the interferometer, and also because the array cannot provide reliable information about seismic waves that are much longer than the diameter of the seismic array (each contributing a $1/f$ at low frequencies). Therefore enlarging the array (without decreasing sensor density in the central part) would increase subtraction performance at the expense of deploying a much larger number of sensors. The same calculation is repeated for the infrasound NN for an infrasound field isotropic over a half space.
\begin{figure}[t]
\includegraphics[width=\columnwidth]{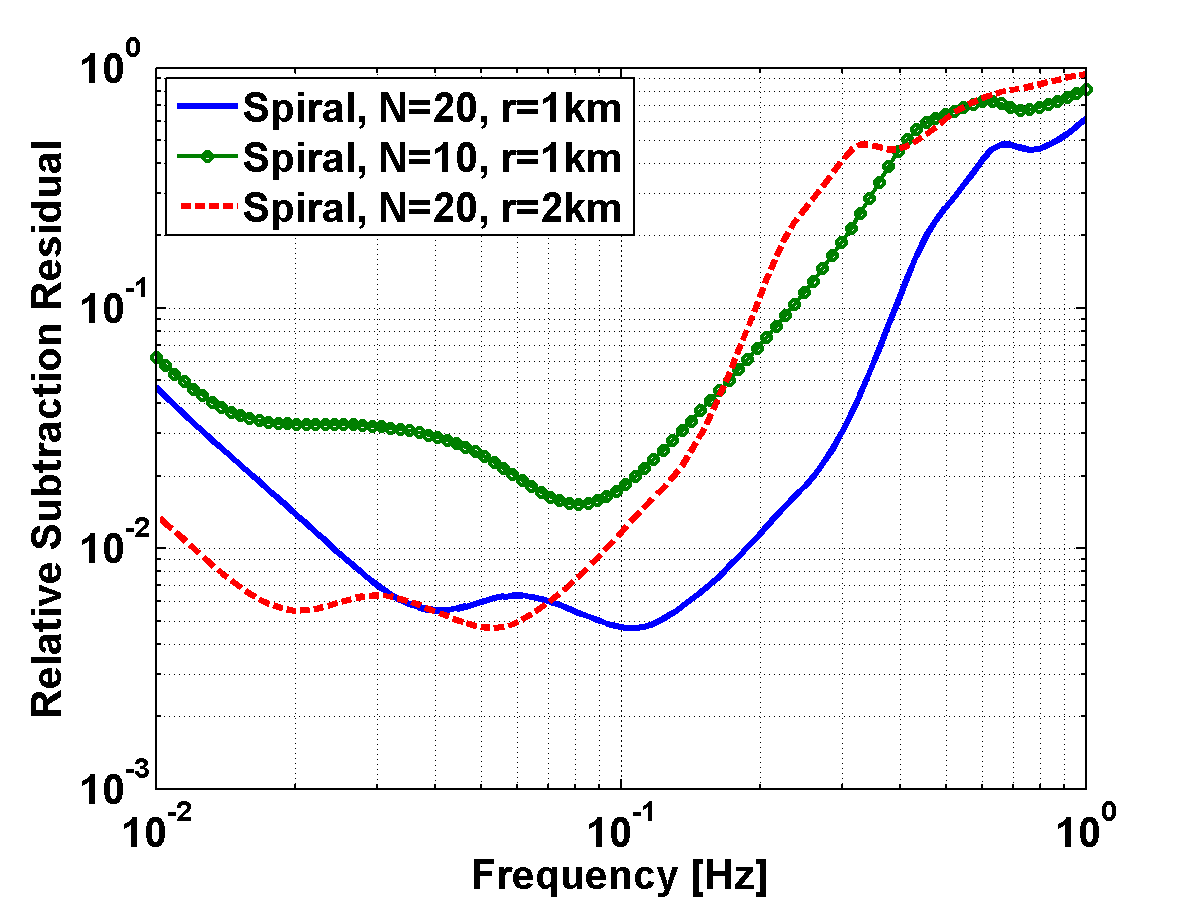}
\caption{Residuals of infrasound gradient NN subtraction for double-wound spiral arrays using microphones with $\rm SNR=1000$. Results are presented for different numbers $N$ of microphones, and different array radii $r$.}
\label{fig:resInfrasound}
\end{figure}
The results are shown in Figure \ref{fig:resInfrasound}. Even though the microphones are assumed to have the same sensitivity to pressure fluctuations as the seismic sensors to ground displacement, less subtraction is achieved. The reason is that NN from sound waves propagating in three dimensions is subtracted using a two-dimensional microphone array deployed on Earth's surface. If it were possible to monitor atmospheric infrasound at different altitudes, then subtraction residuals could be similar to the seismic case. In summary, the results shown in Figures \ref{fig:resRayleigh} and \ref{fig:resInfrasound} demonstrate that it will be very challenging to achieve sufficient NN subtraction. A suppression of the NN by about 4 or 5 orders of magnitude at 0.1\,Hz would be needed to make it comparable to the instrument noise limit. In order to achieve the goal, a larger number of more sensitive sensors will be required, and the arrays should ideally be tailored to the required NN subtraction factors.

We conclude this part with a brief discussion about the sensors required to achieve the NN subtraction goals. The sensitivities of various seismometers was compared in \cite{RiHu2010}. The fact that the seismometer self-noise curves lie a factor 5 or less below the seismic low-noise model at frequencies less than 0.1\,Hz seems discouraging. Also, the best gravimeters (when used as seismometers) barely resolve the global new low-noise model NLNM \cite{Pet1993} at frequencies above 10\,mHz \cite{Goo1999,BaCr1999,FHT1997}, but it is not completely understood what type of noise or environmental couplings are causing sensitivity limits in modern instruments. There are efforts to improve the low-frequency sensitivity of seismometers, and if for example the dominant noise is a result of coupling to the environment, like pressure or temperature changes, then a solution could be coherent noise subtraction using additional thermometers or barometers. The same holds for the gravimeters. Therefore, it is very important to investigate noise in seismometers and gravimeters, otherwise the GW band below 0.1\,Hz may remain inaccessible to ground-based detectors. It has often been proposed to use additional laser interferometers to measure and subtract NN, but these schemes fail since laser interferometers are exclusively sensitive to gravity strains, and it would be impossible to distinguish NN from GWs. Instead, a possible solution would be to sense a degree of freedom of the gravity field that does not have contributions from GWs, but that shows correlations with the strain field with respect to terrestrial perturbations. 

\section{Sources of gravitational waves from 0.01\,Hz to 1\,Hz} 
\label{sec:Sources}

\subsection{Compact binaries}

In this section, we discuss the most well-understood gravitational-wave sources for MANGO, namely compact binaries of white dwarfs, neutrons stars and black holes.  We will first briefly review the evolution of these binaries under gravitational radiation reaction, and then discuss several scenarios in which gravitational waves from these binaries might be detected. 

\subsubsection{Evolution of a compact binary under radiation reaction}

Let us first briefly review the basics of  gravitational waves from binaries in circular orbits (or {\it circular binaries}, which is probably a good approximation in most cases in this frequency range).  This involves both the strain amplitude at a given frequency and the time spent at that frequency, as these both play a role in detectability.  From \citet{1997rggr.conf..447S}, the angle-averaged strain amplitude measured a distance $r$ from a circular binary of masses $m_1$ and $m_2$ (and hence total mass $M\equiv m_1+m_2$ and symmetric mass ratio $\eta\equiv m_1m_2/M^2$) with a binary orbital frequency $f_{\rm bin}$  (and hence gravitational wave frequency $f_{\rm GW}=2f_{\rm bin}$) is
\begin{align}
h=& \displaystyle  {2(4\pi)^{1/3}\over c^4}
\frac{\eta (GM)^{5/3}}{r}f_{\rm GW}^{2/3}
\nonumber\\
= &2.4\cdot 10^{-22}\left[\frac{f_{\rm GW}}{0.01\,{\rm Hz}}\right]^{\frac{2}{3}}
\frac{\eta}{0.25}\left[\frac{M}{2M_\odot}\right]^{\frac{5}{3}}\frac{10\,{\rm kpc}}{r},
\end{align}
where in the second line we normalize to an equal-mass binary ($\eta=0.25$).  Note that for comparable-mass sources,  $\eta$ is close to 0.25; for example, $m_1/m_2=1.5$ gives $\eta=0.24$ and even $m_1/m_2=2$ gives $\eta=2/9$, which is only a $11\%$ change from 0.25. 

From \citet{1964PhRv..136.1224P} the semimajor axis $a$ of a circular binary evolves adiabatically via gravitational radiation as
\begin{equation}
\dot a=-{64 \over 5  c^5}{\eta (GM)^3\over{ a^3}}\; .
\end{equation}
Switching variables to $f_{\rm GW}=(GM/a^3)^{1/2}/\pi$ gives
\begin{equation}
\dot {f}_{\rm GW}={96 \pi^{7/3}\over 5c^5 }
\eta(GM)^{5/3}f_{\rm GW}^{11/3}\; .
\end{equation}
From $\dot f_{\rm GW}$, we can estimate a characteristic time for radiation reaction, 
\begin{equation} 
T\equiv {f_{\rm GW}/{\dot{f}_{\rm GW}}} \,.
\end{equation}
If $f_{\rm GW}$ is at least a factor of a few less than the merger frequency, then the time left before merger, or the additional life time of the inspiral, is 
\begin{align}
\label{tinsp}
&T_{\rm insp}={3\over 8}T  = {5\over{96\pi^{7/3}}}{c^5\over{\eta(GM)^{5/3}}}f_{\rm GW}^{-8/3} \nonumber\\
=&8.2\cdot 10^3\,{\rm yr}\left[\frac{0.25}{\eta}\right]
\left[\frac{M}{2M_\odot}\right]^{-\frac{5}{3}}\left[\frac{f_{\rm GW}}{0.01\,{\rm Hz}}\right]^{-\frac{8}{3}} .
\end{align}

Suppose our detector has a noise spectral density of $S_h$.  Then the signal-to-noise ratio, using the matched-filtering detection technique, is
\begin{equation}
\rho^2 = 4\int_0^{+\infty} df \frac{|\tilde h(f)|^2}{S_h(f)}
\end{equation}
If the detector's spectral density and the amplitude of the GW are both roughly constant, then (as can also be seen using Parseval's theorem)
\begin{equation}
\rho^2 \approx \frac{4 h^2 T}{S_h} \,.
\end{equation}
If $\rho_*$ is the threshold for detectability, then for any GW signal $h$ the detector's maximum spectral density is given by
\begin{equation}
\label{sstar}
S_* = \frac{4}{\rho_*^2} h^2 \min(T_{\rm insp},T_{\rm obs})
\end{equation}
Here we have taken the minimum of the inspiral time $T_{\rm insp}$ and the observation time $T_{\rm obs}$. 


\subsubsection{Individual Neutron-star binaries}

\begin{figure}[htb]
\begin{center}
\includegraphics[scale=0.45]{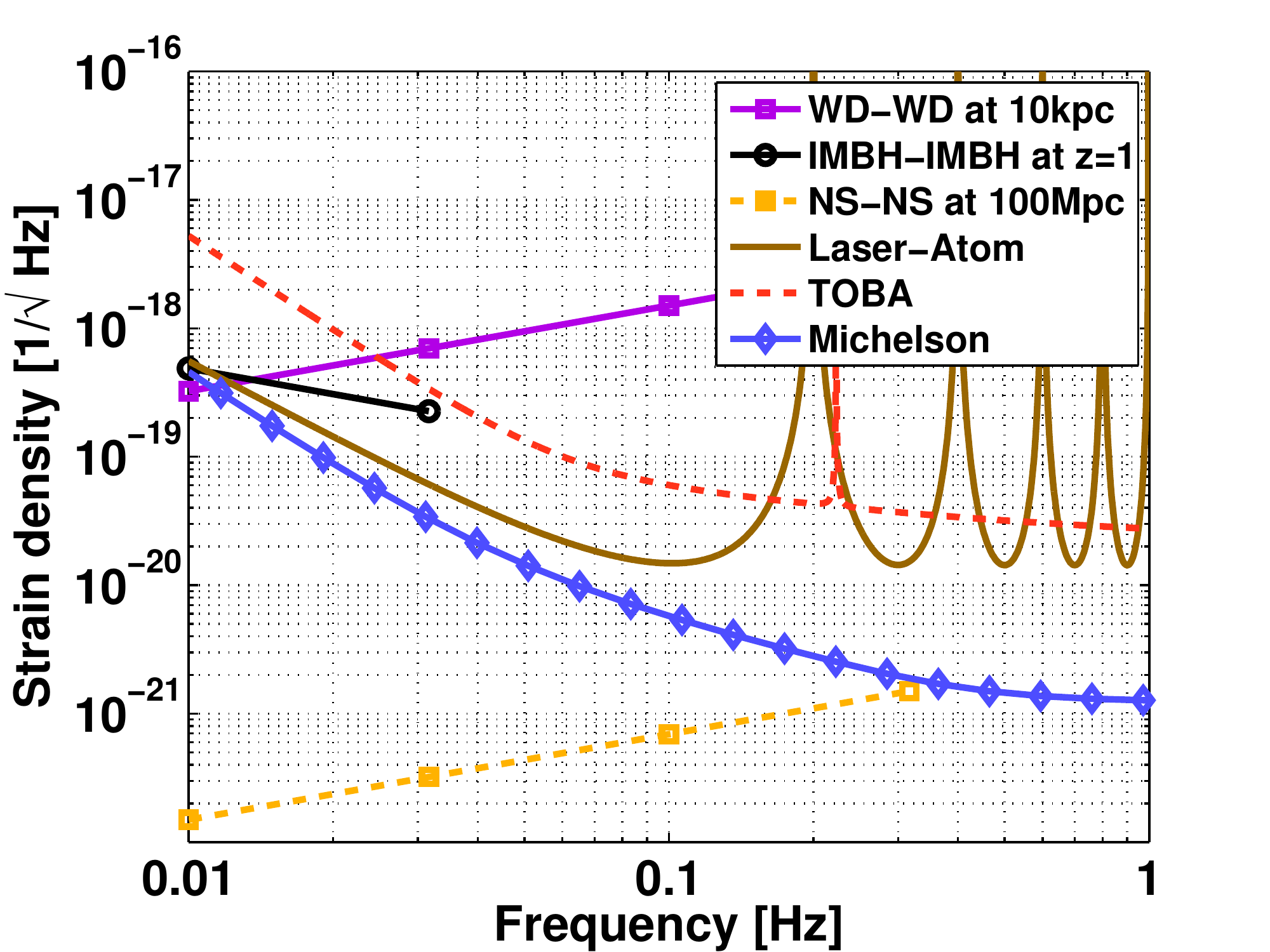}
\caption{Example tracks with residence-time-weighted spectral density for a double white dwarf merger at 10\,kpc (solid red line; both have a mass of $0.6~M_\odot$), a double neutron star merger at  100\,Mpc (black dotted line; both have a mass of $1.5~M_\odot$), and a double intermediate-mass black hole merger at $z=1$ (blue dashed line; both have a mass of $10^4~M_\odot$).  The IMBH-IMBH curve is terminated at the ISCO (innermost stable circular orbit) frequency as we see it in our frame (0.1\,Hz).}
\label{fig:flowchart}
\end{center}
\end{figure}

From Eq.~\eqref{tinsp}, we see that for white dwarf and neutron star binaries below $f_{\rm GW}\sim 0.1$\,Hz, the inspiral time is greater than $\sim 10^7$\,s, which we use as a fiducial observation time for MANGO, hence for those sources $10^7$\,s is the relevant time in Eq.~\eqref{sstar}. For more massive sources, such as IMBH-IMBH binaries, the inspiral time is relevant because it is shorter than $10^7$\,s.  Assuming a distance of 10\,kpc for WD binaries, 100\,Mpc for NS binaries, and $z=1$ for IMBH binaries, and assuming $\rho_*^2 = 4$, we plot in Fig.~\ref{fig:flowchart} the minimum spectrum for the detector at different frequencies.

In order to estimate the relevant distance for each type of binary event, we have to use our knowledge about their rates.  Suppose the rate per Milky-Way Equivalent Galaxy (MWEG) is $\mathcal{R}_{\rm MWEG}$.  Then for a frequency $f_{\rm GW}$ and a corresponding lifetime of $T_{\rm insp}$, the probability that there is at least one such binary in a MWEG is
\begin{equation}
p =1 -\exp(-\mathcal{R}_{\rm MWEG} T_{\rm insp}) \approx \mathcal{R}_{\rm MWEG} T_{\rm insp}\,
\end{equation}
where in the last expression we assume $\mathcal{R}_{\rm MWEG} T_{\rm insp} \ll 1$.

For two neutron stars of 1.4\,$M_\odot$, and assuming a galaxy rate from  1 -- 1000 $\mbox{Myr}^{-1}$~\cite{AbEA2010}, we have a probability, ranging from 0.4\% to 98\%, to have at least one binary neutron star with a gravitational-wave frequency at or above 0.01\,Hz.  For the most likely rate of 100 $\mbox{Myr}^{-1}$, that probability becomes 34\%. From Fig.~\ref{fig:flowchart}, it is plausible for MANGO to reach 10\,kpc, and therefore the chance for MANGO to detect a neutron star binary in our galaxy is non-negligible. 

One has to reach substantially farther in order to detect binaries from other galaxies.   In the most pessimistic case, we will have to reach $\sim 700$\,MWEG in order to guarantee a binary with 95\% confidence, which implies a horizon distance of 56\,Mpc.  If we use the most likely rate, then in order to guarantee the same likelihood of detection we need to reach around 7.2\,MWEGs, which is achievable if we can reach a  horizon distance of $\sim 5$\,Mpc.  Here we have used the conversion formula of 
\begin{equation}
\label{eqng}
N_G =\frac{4}{3}\pi \left(\frac{D_{\rm horizon}}{\rm Mpc}\right)^3 (2.26)^{-3}  0.0116
\end{equation}
which is Eq.~(4) of Ref.~\cite{AbEA2010} when $D_{\rm horizon}$ is larger than $\sim 30$\,Mpc, and Fig.~1 of the same paper for smaller distances. Noting that $D$ in Eq.~\eqref{sstar} is the distance reachable by a detector after averaging, instead of the horizon distance, with a conversion factor of
\begin{equation}
D =\frac{D_{\rm horizon}}{2.26}
\end{equation} 
This means, at the most likely rate, we need $D \approx 2.2\,$Mpc, while in the least rate, we need $D\approx 25 \,$ Mpc.  If we account for additional integration time, then from Fig.~\ref{fig:flowchart}, it is plausible that MANGO can detect a NS binary from nearby galaxies.  

 The rate per galaxy of BH-NS and stellar-mass BH-BH mergers is less than for NS-NS, and the binaries are more massive and thus spend less time above 0.01\,Hz, so it is highly unlikely that such binaries are in the MANGO band in the Galaxy at the moment.  Given that even the existence of intermediate-mass black holes is still under debate, estimates of their rates are even less certain.  However, if multiple IMBHs can form in a dense stellar cluster (e.g., \citet{GuEA2006}) or separate clusters with IMBHs merge (e.g., \citet{ASFr2006}, \citet{ASEA2012}), then depending on the fraction of clusters that form IMBHs and their masses and merger efficiencies there could be tens of mergers per year visible out to the $z\sim 1$ range of MANGO (\citet{FrEA2006}, \citet{MaEA2008}, \citet{GaEA2011}).


\subsubsection{Individual White-Dwarf Binaries}


Let us estimate the galactic merger rate of WD binaries.  Collisions between two white dwarfs with a combined mass greater than the $\sim 1.4~M_\odot$ Chandrasekhar mass are candidates for Type~Ia supernovae, and even collisions between two typical white dwarfs of mass $0.6~M_\odot$ release $\sim 10^{50}\,{\rm erg}$ in gravitational binding energy, so we would expect them to be easily detectable.  Thus a rate that implies such occurrences in our Galaxy more than once every few years is not plausible; we would have seen them.  We note, however, that for two $0.6~M_\odot$ white dwarfs the inspiral time from 0.01~Hz is $T_{\rm insp} \sim 2\times 10^4$\,yr, so if there are currently a few hundred such binaries in the Galaxy their merger rate would only be one per few decades (note that the expected number of binaries is $N =\mathcal{R} T_{\rm insp}$) which could have been missed or misidentified.  If there are supposed to be a few thousand, however, the rates get too high to miss. In other words, such a rough estimate puts 
\begin{equation}
\mathcal{R}_{\rm WD} \stackrel{<}{_{\sim}} 10^{-2}/{\rm yr},\quad N \stackrel{<}{_{\sim}} 10^2\,.
\end{equation}

Other considerations lead to more concrete estimates. For example, from observations of WD binaries, Badenes and Maoz~\cite{Badenes.2012} estimate that in our galaxy, the merger rate of super Chandrasekhar WD binaries (i.e., those with total mass greater than the Chandrasekhar mass) is $\sim 6.4\times 10^{-4}$/yr, and the total rate is $\sim 8\times 10^{-3}$/yr\,.  For a Chandrasekhar WD binary, its inspiral time at 0.01\,Hz is $1.5\times 10^4\,$yr.  Noting that lifetime at a fixed frequency increases with decreasing total mass, we can estimate that 
\begin{equation}
N_{M>1.4M_\odot} \sim 10\,,\quad N_{M<1.4M_\odot} \sim 100\,.
\end{equation}
This is compatible with the rough estimate above, and suggests that a MANGO like detector would have tens of sources in band. The lowest-period known WD/WD binary is HM Cnc with a corresponding GW frequency of 6\,mHz \cite{RoEA2010}. Therefore, all known binaries lie below the MANGO band.





\subsubsection{Stochastic Background from Galactic and Extragalactic Binaries}


If a large population of binaries with unknown parameters is viewed collectively as a source, the gravitational waves it emits may be viewed as a ``stochastic background'' \cite{ReMa2008,Reg2011}. However, in some cases, one can estimate parameters of some of the (stronger) binaries, and ``resolve'' part of this ``stochastic background'' into a complex but deterministic waveform \cite{CuHa2006,HaEA2008}.

Our ability to estimate parameters of the binary depends on two factors: (i) the duration of the observation and the individual waves and (ii) the sensitivity of our detector.  Let us try to understand this for the simplest case in which all binary waves are quasi-monochromatic but with a finite lifetime $\tau_*$.  Let us first select a finite subpopulation that already contributes to most of the spectrum;  for example, a large enough finite cut-off distance. In this case, we first require (i) the lifetime of each wave $\tau_*$ and the observation time $T_{\rm obs}$ must both be long enough, so that within each frequency bin with a bandwidth of $2/\min(\tau_*,T_{\rm obs})$, there is {\it at most one} binary of the subpopulation; we then require that (ii) in those bins with a binary, our detector to have high enough sensitivity to detect the wave emitted by the binary.

As discussed by \citet{2003MNRAS.346.1197F}, for a 1-year observation, the galactic population of WD-WD binaries above 0.01\,Hz (in the MANGO band) are all individually resolvable.  On the other hand, the number of all extragalactic binaries is larger by a factor equal to the number of MWEGs in the universe, i.e., $10^{10}$ or more, and hence forms a population that is only individually resolvable at frequencies too high for most white dwarfs to reach, i.e., above $\sim 0.1$\,Hz, giving rise to a stochastic background below 0.1\,Hz (see their Figure~17).  \citet{2003MNRAS.346.1197F} also give their expected extragalactic WD-WD background strength in their Figure~16.  They express it as $\Omega_{\rm gw}(f)$, which is the energy density in a frequency band of width $f$ centered on $f$, expressed as a fraction of the critical energy density for the universe.  From equation~(5) of \citet{Phi2001}, the characteristic strain amplitude is related to $\Omega_{\rm gw}(f)$ by
\begin{equation}
\label{hc}
\begin{array}{rl}
h_c(f)&=\left[4G\rho_c\Omega_{\rm gw}(f)/f^2\right]^{1/2}\\
&\approx 1.6\times 10^{-22}(\Omega_{\rm gw}/10^{-12})^{1/2}
(0.01~{\rm Hz}/f_{\rm GW})\; ,\\
\end{array}
\end{equation}
where in the second line we have substituted $\rho_c=9.5\times 10^{-30}~{\rm g~cm}^{-3}$ (valid for a Hubble constant $H_0=71~{\rm km~s}^{-1}~{\rm Mpc}^{-1}$). On the other hand, using two co-located detectors with noise spectral density $S_h$, the characteristic $h$ one can detect, after a duration of $T_{\rm obs}$, is
\begin{align}
\label{hcdetect}
h_c^*(f) \approx & (2\pi f T_{\rm obs})^{1/4}\,\sqrt{f S_h(f)}  \nonumber\\
\approx & 2\cdot 10^{-22} \left[\frac{S_h^{1/2}(f)}{10^{-19}\,{\rm Hz}^{-\frac{1}{2}}}\right] \left[\frac{f}{0.01\,{\rm Hz}}\right]^{\frac{3}{4}} \left[\frac{T}{10^8\,{\rm s}}\right]^{\frac{1}{4}}
\end{align}
[Note that when there are many sources, we can integrate over the entire observation time of $10^7$\,s -- $10^8$\,s without regard to how long each individual source takes to spiral in.]

Figure~16 of \citet{2003MNRAS.346.1197F} suggests that $\Omega_{\rm gw}$ may peak at $\sim 10^{-11}$ around 0.01\,Hz. According to Eqs.~\eqref{hc} and \eqref{hcdetect}, this background could be detectable with a $10^8$\,s MANGO integration but not likely with $10^7$. Above 0.01~Hz the expected background becomes progressively more uncertain, but possibilities exist up to $\sim 0.1$\,Hz (after which the number of sources falls rapidly).

In comparison, stochastic background from other extragalactic binaries (e.g., NSNS) are much less in magnitude, and will be buried under the background arising from extragalactic WD binaries. 

\subsection{Helioseismic and Other Pulsation Modes}

Given that the Sun is extremely close and has strong helioseismic $p$-modes at around $\sim 300$\,s~\cite{ThEA2007}, one might hope that gravitational radiation from these modes could be seen with MANGO. Also the solar $g$-modes, which have not been definitively detected yet, could potentially be observed. Their periods are all greater than about 45\,min so that the detector would lie in the near-zone gravitational field \cite{ApEA2010}.

Ref.~\citet{1996PhRvD..54.1287C} addresses this exact problem for LISA (which is more sensitive than MANGO at the relevant frequencies) and concludes that unless there are orders of magnitude more energy in the modes than expected, they will not be detectable even at 1\,AU.  This allows us to conclude that, unless other stars (particularly M dwarfs, whose general frequencies should be higher than those of the Sun) have vastly greater energy in the modes than the Sun does, the cumulative gravitational wave background from their pulsations is much less than the amplitude from the Sun. This is for the same reason that the Sun has a larger optical flux than all other stars combined; the gravitational wave energy will add incoherently.  Thus stellar pulsations will not be detectable.

\subsection{Supernovae}
It is very unlikely that GW emission from supernovae would be detected with MANGO. However, computational models of supernovae are not converged yet, and each improvement in numerical technology (three dimensional, fully general relativistic, better neutrino transport, inclusion of rotation, etc.) has brought surprises \cite{OtEA2013}. It is expected that most of the gravitational wave power will be at frequencies $>100$\,Hz (see, e.g., Figure~4 of \citet{2009CQGra..26t4015O}).  Thus these are not likely to be detectable $<1$\,Hz gravitational wave sources even if a supernova happened in our Galaxy, but we should keep an open mind; given that simulations cannot be run for many seconds, perhaps there are unsuspected modes at a few tenths of a Hz.

\subsection{Primordial stochastic background}
It has been argued that the most important gravitational wave detection would be one from the very early universe (e.g., from the inflationary era) because this would give us information from an otherwise highly opaque epoch.  Primordial stochastic backgrounds are predicted among others by inflationary, pre-Big-Bang and cosmic string models \cite{AbEA2009}. In standard inflationary models, however, the strength of this signal is tiny.  From, e.g., Figure~3 of \citet{2003gr.qc.....3085B} one finds that $\Omega_{\rm GW}$ from the inflationary gravitational wave background is likely to be less than $\sim 10^{-15}$, which is not only far below what MANGO could detect, but as we discussed earlier it will be completely masked by the unresolvable extragalactic WD-WD foreground up to $\sim 0.1$\,Hz.  It has been suggested that the $\sim 0.1$\,Hz and above region will be ``clean" in the sense that all foreground (i.e., redshifts in the single digits!) sources will be individually resolvable, hence any sufficiently sensitive instrument might detect the inflationary background \cite{CuHa2006}.  This presupposes, however, that the foreground sources can be subtracted with extreme fidelity, on the order of a part per thousand or possibly much better. Thus this seems unlikely.  There is always the possibility of a surprise source in just the right frequency band, and Big Bang nucleosynthetic constraints are only on the order of $\Omega_{\rm GW}<10^{-5}$ so a detectable stochastic background is possible, but it is not a probable source. The most sensitive searches of stochastic backgrounds require a network of detectors to search for correlations between detectors. In general, these searches are based on the assumption that no other correlated effects occur. However, several possible environmental influences have been identified that could produce correlated noise in two detectors separated by a large distance such as the Schumann resonances \cite{TCS2013}. It can be expected that this problem is more significant at lower frequencies, and a careful analysis should be carried out.

\section{Conclusions}
\label{sec:Conclusion}
We have described three potential detectors in the 0.1 -- 10\,Hz band which can be astrophysically 
interesting and which would be complementary to the audio frequency GW detectors 
(LIGO, GEO, Virgo, and KAGRA) as well as the space based detectors such as eLISA and DECIGO.

The key to this possibility is that the strain sensitivity in this band can be orders of magnitude 
worse than in the audio band, due to the fact that (i) the strain amplitudes are larger and (ii) 
that the sources are much longer lived.

So far, the best strain sensitivity at 0.1\,Hz is $10^{-8}\,\rm /\sqrt{Hz}$, achieved with a prototype 
TOBA~\cite{IsEA2011}. Significant experimental challenges must be overcome in order to make 
any of these types detectors a reality. However, the added astrophysics which can be done with 
these instruments demands that we take on the challenge.

\acknowledgments
We thank S.~Nissanke for helpful discussions about low-frequency GW sources. 
The research of YC is supported by NSF Grant PHY-1068881 and CAREER Grant 
PHY-0956189, as well as the David and Barbara Groce Fund at Caltech. ME and RXA 
are supported by the National Science Foundation under grant PHY-0757058. The 
research of MCM was supported in part by a grant from the Simons Foundation 
(grant number 230349). MCM also thanks the Department of Physics and Astronomy 
at Johns Hopkins University for their hospitality during his sabbatical. BS acknowledges 
the support of the Australian Research Council. A part of this work was supported by 
JSPS KAKENHI Grant Number 90313197. This manuscript has the LIGO document number 
LIGO-P1300100. The research of JH for this paper was carried out at Caltech. HM thanks the National Science Foundation, the National Aeronautics and Space Agency, and the David and Lucile Packard Foundation.

\raggedright
\bibliography{references}

\end{document}